\documentclass[a4paper,fleqn]{cas-dc}

\usepackage[authoryear,longnamesfirst]{natbib}
\usepackage{subcaption}
\def\tsc#1{\csdef{#1}{\textsc{\lowercase{#1}}\xspace}}
\tsc{SIW} %Definition of single impact wave
\tsc{ATM} %make a definition for the Atmosphere

\usepackage{siunitx}

\newcommand{\eref}[1]{Eq.~\eqref{#1}}

\newcommand{\fref}[1]{Fig.~\ref{#1}}
\newcommand{\frefs}[1]{Figs.~\ref{#1}}
\newcommand{\sref}[1]{Sec.~\ref{#1}}
\newcommand{\tref}[1]{Tab.~\ref{#1}}

\newcommand{\cref}[1]{Chapter ~\ref{#1}}

\newcommand{\magenta}[1]{\textcolor{black}{#1}}

\DeclareSIUnit\bar{bar}

\begin{document}
\let\WriteBookmarks\relax
\def\floatpagepagefraction{1}
\def\textpagefraction{.001}

\shorttitle{Predicting GP dynamics using ML techniques}

\shortauthors{Ezeta \& D\"uz}

\title [mode = title]{Predicting the dynamics of a gas pocket during \magenta{breaking} wave impacts using machine learning}                      

\author{Rodrigo Ezeta}[type=editor,
                        auid=000,bioid=1,
                        prefix=,
                        role=,
                        orcid=0000-0001-5366-6920]

\cormark[1]

\ead{r.ezeta@marin.nl}

\ead[url]{https://www.marin.nl/en}

\credit{Conceptualization, Data curation, Formal analysis, Investigation, Methodology, Visualization, Writing – review and editing}

\affiliation{organization={Maritime Research Institute Netherlands (MARIN)},
    addressline={Haagsteeg 2}, 
    city={Wageningen},
    % citysep={}, % Uncomment if no comma needed between city and postcode
    postcode={6708 PM}, 
    country={The Netherlands}}

\author{B\"ulent D\"uz}[type=editor,
                        auid=000,bioid=2,
                        prefix=,
                        role=,
                        orcid=0000-0002-0885-281X]
%\ead{b.duz@marin.nl}
%\ead[URL]{https://www.marin.nl/en}

\credit{Conceptualization, Data curation, Formal analysis, Investigation, Methodology, Visualization, Writing – review and editing}

\cortext[cor1]{Corresponding author}

\begin{abstract}
We investigate the feasibility and accuracy of a machine learning model to predict the dynamics of a gas pocket that is formed when a breaking wave impacts on a solid wall. The proposed ML model is based on the convolutional long short-term memory structure and is trained with experimental data. In particular, it takes as input two high-speed camera snapshots before impact and produces as output six scalars that describe the dynamics of the gas pocket. The experiments are performed in a wave flume, where we use solitons -- in combination with a bathymetry profile -- to generate wave breaking close to a solid wall which is instrumented with dynamic pressure sensors. By varying the water depth $h_\ell$ and the parameter $\alpha = A/h_\ell$,  \magenta{where $A$ is the soliton wave amplitude}, we are able to generate a family of unique breaking waves with different gas pocket sizes and wave kinematics. In this so-called phase space of wave generation ($h_\ell$, $\alpha$), we perform experiments on 67 different wave states that form our dataset. Experimentally, we find that the frequency of oscillation of the gas pocket can be attributed to the initial volume of gas plus a geometric correction and that the maximum and minimum pressures are qualitatively well captured by the one-dimensional Bagnold model. In terms of the ML model, we compare its performance to the experimental data and find that the model quantitatively reproduces the trends found in the experiments -- in particular for the maximum and minimum pressure in the gas pocket and the frequency of oscillation. 
\end{abstract}

\begin{highlights}
\item We conduct a dedicated experimental campaign to generate a family of breaking waves with different amplitudes and kinematics in a wave flume . In our experiments, a breaking wave -- which is generated when a solitary wave interacts with a bathymetry profile -- impacts onto a solid wall leading to the formation of a gas pocket.
\item We train a machine learning model based on the convolutional long short-term memory structure whose input consists of two high-speed camera snapshots (before and upon impact) in order to predict six scalars that describe the dynamics of the gas pocket oscillations.
\item The proposed machine learning method is able to reproduce the trends found in the experiments -- in particular for the maximum and minimum pressure in the gas pocket and the dominant frequency of oscillation.
\end{highlights}

% Keywords

\maketitle

\section{Introduction}

When a gravity wave of a certain size, shape and velocity impacts onto a structure, a spatio-temporal force is exerted onto it. From an industrial point of view, the knowledge of the magnitude and the duration of this loading is of extreme relevance as it can determine the adequate design loads. These loads -- along with an ample safety margin -- are then taken into account when designing the structure in order to guarantee its integrity throughout a long period of time \citep{cuomo2011, vanessen2023}. This type of methodology is ubiquitous in industry and is used in many applications where the structure (i.e. ships, breakwaters, storm surge barriers, wave energy converters, offshore wind turbines, containment systems for liquid fuel, etc.) is expected to be exposed to a variety of wave impacts throughout its lifetime \citep{peregrine2003,faltinsen2004, faltinsen2009,kapsenberg2011,dias2018}. In practice however, these loads are challenging (if not impossible) to obtain at full scale. Thus, the state-of-the-art methodologies rely on model tests where the structure of interest is exposed to wave impacts in laboratory conditions. Here, dynamic pressure (or force) sensors are installed on the structure and impact pressures (or loads) at this scale are subsequently upscaled -- typically using Froude scaling.

At model scale, a common laboratory practice is to use irregular waves (i.e. various frequency components and amplitude) which are usually generated via a certain empirical wave spectrum. The waves are designed in such a way that they break near the model of interest. As these breaking events are random in nature and the tests are executed for several hours, this yields a large statistical ensemble of wave impact pressures \citep{vanessen2023b,scharnke2023}. While this approach may be useful to calculate global statistical properties (i.e. exceedance probabilities) due to inertial effects, it typically leaves out local information such as the influence of surface tension \citep{stagonas2011,fortin2020,erinin2023}, gas-to-liquid density ratio (DR) effects \citep{etienne2018} and free-surface instabilities \citep{karimi2015, vanmeerkerk2020, vandermeer2022} on the global wave shape (GWS). \magenta{In particular, we highlight the role of free-surface instabilities which seemingly tend to ``redistribute'' the impact forces leading to a reduction of the overall impact pressures \citep{maillard2009, fortin2020,vandermeer2022}. This is highly relevant for the transport of cryogenic fuels such as LNG or LH2. The origins of these Kelvin-Helmholtz instabilities are up to this date an active topic of research where single impact waves (SIWs) are often used to isolate these local effects.}

A common practice to generate SIWs in wave flumes is to use focusing waves \citep{chan1994,hofland2011,scharnke2023}. Here, the dispersion of single component gravity waves is used to ``focus'' all the wave energy at a single point in time and space. When the location of this ``focusing point'' -- as it is commonly referred to -- is set close to the structure, wave breaking occurs provided that the amplitude of the wavelet surpasses that of the breaking criterion. When the structure is a solid wall, the breaking wave can in turn entrap gas which forms a gas pocket (GP). \magenta{Consequently,} the gas pocket experiences a series of compression and expansion cycles that results in a damped oscillating forcing on the wall. Depending on how far the focusing point is set away from the wall, one can obtain different GWSs upon impact (i.e. gentle slosh, flip through, breaking wave, plunging wave) \citep{hofland2011}; and consequently, one can obtain different gas pocket sizes. In addition to wave focusing, solitary waves (or solitons) can also be used to generate SIWs. Here, however, a beach needs to be installed close to the wall as otherwise the solitary wave would not break \citep{kimmoun2009,xiang2020}.

Independently of the type of wave that is used to generate breaking waves, the crux of the problem is then to obtain precise measurements of the wave impact pressures (or forces). What we mean by ``precise'' here refers to pressure/force sensors that are accurate enough to resolve both the time and pressure scales of the impact and sufficiently small so as not introduce biases due to the sensor size \citep{bogaert2018}. Thus, the main question we pose in this work is the following: Can a machine learning (ML) model trained on experimental data predict the impact pressures of a SIW given only ``visual'' information from the GWS?

Machine learning applications in the maritime domain are gaining popularity as evidenced by many works found in the literature. In the context of wave breaking, researchers have mainly focused on its detection and classification. \magenta{Traditionally, classification of breaking waves has been studied both experimentally and numerically, where criteria have been developed based on physical or environmental quantities. Examples of these include the velocities at the wave crest, wave height, slope of the beach, water depth, etc. In contrast, in the ML-based approaches, one asks the ML model to extract the necessary information\textbackslash feature from the data itself.} In \citet{liu2024} for instance, a ML model is developed using simulation data to capture breaking in focused waves, random waves and modulated plane waves. In \citet{eadi2021}, a method based on convolutional neural networks (CNNs) is developed to detect active wave breaking in video imagery data. Here, various data sources are used including real world data for which the model shows a promising performance. In \citet{duong2023}, a multi-layer perceptron (MLP) is adopted to predict the breaker height using experimental data, where wave breaking occurs due to a varying bathymetry in the experiments. In \citet{buscombe2019}, CNNs are adopted to estimate wave breaking type from close-range monochrome infrared imagery of the surf zone. Here, observations of breaking waves in the outer surf zone are collected using thermal infrared imagery during a field campaign. \citet{smith2023} worked on the same problem using the same data as in \citet{buscombe2019}. However, they adopted CNNs as basic feature extractors and a classifier was then trained on top of them in order to classify images of non-breaking, plunging and spilling breaking waves. In \citet{yun2022}, a multi-layer perceptron is used to estimate the breaking wave height and wave-breaking location from the input of the bottom slope, deep-water wave height and wave period. In \citet{tu2018}, a detection of plunging breaking waves is formulated as a binary classification, where a logistic regression algorithm is used together with experimental data from a wave flume. 

Similarly, ML models are applied to predict the impact of water waves on structures and shorelines. For instance, in \citet{lay2021}, a system utilizing video camera observations is developed to monitor waves impacting urban shorelines. Wave run-up on beaches are studied in \citet{kim2024, saviz2024, tarwidi2023}, where various ML techniques such as the XGBoost and the conditional generative adversarial network are used. Furthermore, in \citet{pena2021} a ML model is developed to predict three-dimensional nonlinear wave loads and the run-up on a fixed structure using data from Computational Fluid Dynamics simulations.

In the present study, we train a ML model in order to predict the impact pressures that are generated by a SIW onto a solid wall. In particular, we focus on the gas pocket that is generated when the impacting wave entraps a non-condensable gas, i.e. air at ambient conditions. We choose to restrict the study to the gas pocket dynamics as the impact pressures are highly reproducible in the experiments. This is in contrast to the impact pressures associated to the wave crest or the building jets where a large variability is present \citep{bogaert2018,ezeta2023}. The proposed ML model is named as cLSTM as it is based on the convolutional long short-term memory structure. This model takes two high-speed camera snapshots (before and at impact) as input and outputs the dynamics of the gas pocket. More concretely, we investigate whether the cLSTM model can predict the dynamics of the gas pocket in terms of the max. and min. pressure, the times at which these values are reached, the dominant frequency of oscillation as well as the positive and negative decay rates. The cLSTM model is trained with a family of breaking waves with different gas pocket sizes and kinematics. Similar to \citet{kimmoun2009}, the breaking waves in our study are generated when a soliton interacts with a beach installed close to the solid wall.

The paper is organized as follows. In \cref{chap:materials_exp}, we describe the experimental set-up, provide a comprehensive description of the wave generation and define the scalars that we use to quantify the gas pocket dynamics --  these scalars are ultimately the output of the ML model. In \cref{chap:materials_ml}, we provide the details of the ML model. In \cref{chap:results_exp}, we first look at the results of the experiments and discuss their behavior as a function of the wave generation parameters. In particular, we compare the experimental dominant frequency to the theoretical estimate of \citet{topliss1992}. Additionally, we compare the max. and min. impact pressures with the well-known one-dimensional Bagnold model \citep{bagnold1939} to elucidate some of the trends. Next, in \cref{chap:results_ml}, we discuss the performance of the cLSTM model and compare it to a MLP baseline. In addition, we compare both models to experimental data and evaluate the errors, i.e. the difference in percentage between the predictions and the experimental data. Finally, in \cref{chap:conclusions}, we present some conclusions and future work.

\section{Materials and Methods: Experiments}
\label{chap:materials_exp}

\subsection{The Atmosphere facility (ATM)}
\begin{figure}[ht]
    \centering
    \includegraphics[width = \linewidth]{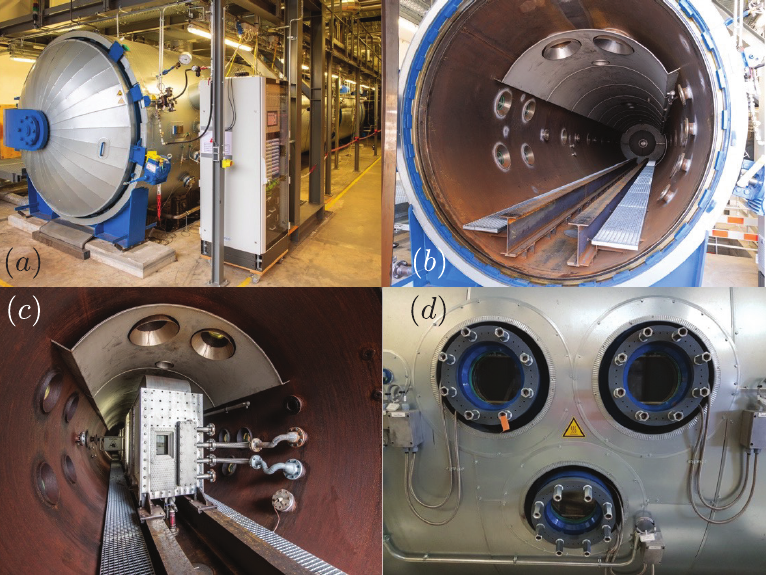}
    \caption{Photographs of the experimental set-up. (a) The Atmosphere (ATM) facility. (b) The autoclave of the ATM (without the flume). (c) The flume inside the autoclave. (d) Three observation windows on the autoclave close to the impact wall where the high-speed cameras are typically installed.}
    \label{fig:atm_photos}
\end{figure}

The experiments are performed in the Atmosphere facility (ATM) at MARIN. The ATM (\fref{fig:atm_photos}a) is a large-scale experimental facility where waves generated in a flume can be exposed to extreme environmental conditions. The ATM consists of a cylindrical autoclave $\qty{15}{m}$ in length and $\qty{2.5}{m}$ in diameter (\fref{fig:atm_photos}b). The novelty of this state-of-the-art facility is that, inside of the autoclave, three  parameters can be independently controlled and monitored. Namely, the ullage pressure $p_u$ in the range $p_u\in [0.02,10] \ \text{bar} $, both the gas $T_g$ and liquid $T_l$ temperature in the range $T_\ell,T_g \in[15, 200] \ \unit{\celsius}$ and the molar mass $M$ of the gaseous phase in the range $M\in[4, 39.95] \times 10^{-3} \ \unit{\kilogram\per\mol}$. This is achieved by using gas mixtures of either non-condensable gases (Helium He, Nitrogen $\text{N}_2$, air, Argon $A_r$) or condensable gases (water vapor). In the present study however, all experiments are performed at ambient conditions, i.e. $T\equiv T_g =T_\ell \approx \qty{20}{\celsius}$ and $p_u \approx \qty{1}{\bar}$, with air as the working gas, i.e. $M=\qty{28.97}{\kilogram \per \mol}$. \magenta{In this configuration, the speed of sound in the gas phase remains relatively constant at $c_g=\sqrt{\gamma R T/M} \approx \qty{343}{\meter \per \second}$, where $\gamma=1.4$ is the adiabatic constant for air. }

\begin{figure*}[ht]
    \centering
    \includegraphics[width = \textwidth]{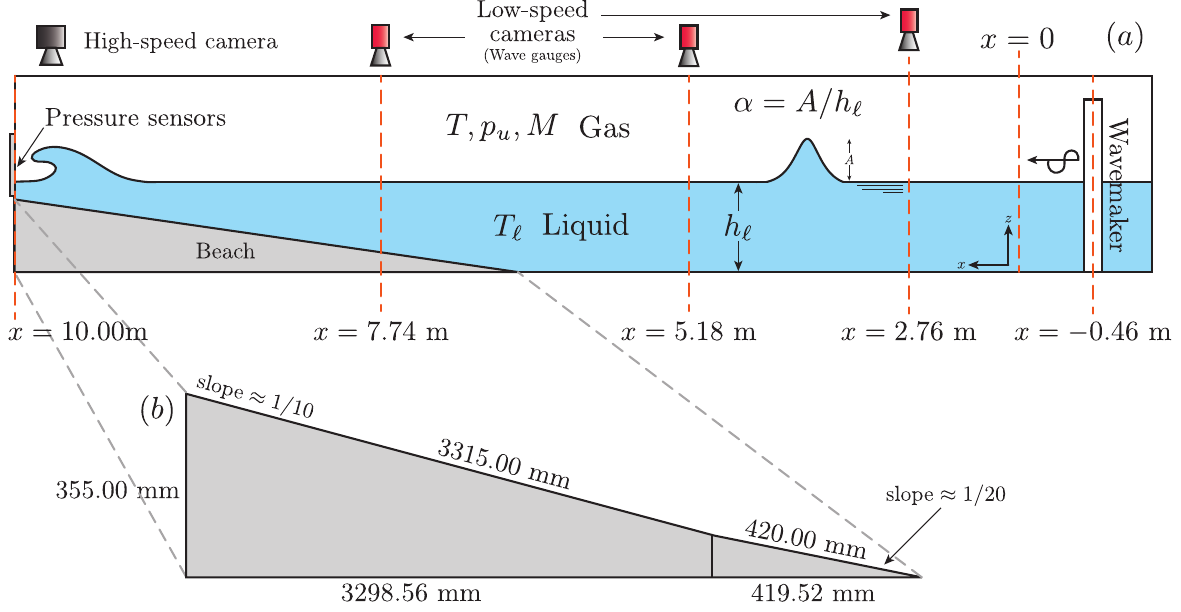}
    \caption{(a) Sketch of the flume inside the ATM along with some relevant instrumentation used in this work (not to scale): wavemaker, 3 low-speed cameras which serve as wave gauges, one high-speed camera at the impact wall, pressure sensors and the beach. (b) The dimensions of the beach. }
    \label{fig:sketch_flume}
\end{figure*}

Inside of the autoclave lies a flume (\fref{fig:atm_photos}c) with length $L=\qty{12.64}{\meter}$ and width $W=\qty{0.6}{\meter}$ and a piston-type wavemaker (WM) which we use to generate waves. Additionally, the ATM has 17 windows (\fref{fig:atm_photos}d) that are used for observation and 12 feedthroughs that can be used to accommodate additional instrumentation. Further instrumentation of the ATM includes an encoder at the WM that measures its motion as a function of time, four \textit{Manta G235-B} -- which we use as wave gauges. These cameras are called ``low-speed'' cameras and operate at  200 fps. In addition, the facility has two \textit{Photron SAX} high-speed (HS) cameras  we operate in this work at 4000 fps. These cameras allow us to capture the wave impacts in detail. In \fref{fig:sketch_flume}a, we show a sketch of the flume along with some of the instrumentation and the relevant fluid control parameters. For more information about the ATM facility, we refer the reader to \citet{novakovic2020} and \citet{ezeta2023}.

\begin{figure*}[ht]
    \centering
    \includegraphics{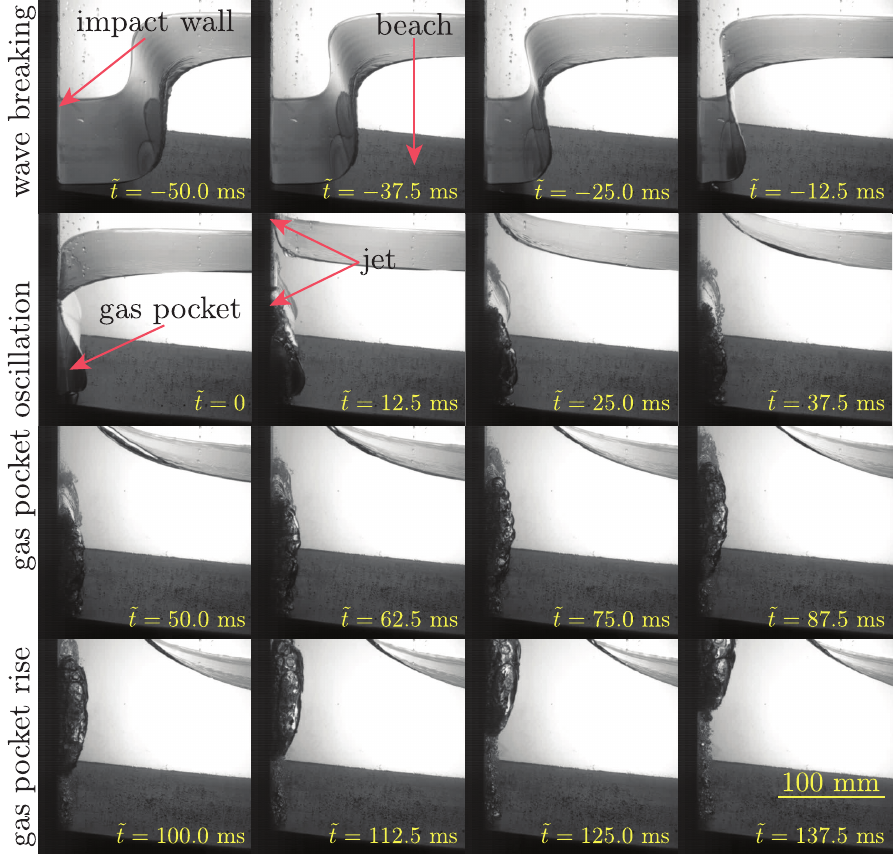}
    \caption{An example of a SIW in the ATM. This breaking wave is generated when a soliton interacts with the beach as described in \sref{sec:wave_generation}. This SIW corresponds to the wave state $h_\ell = \qty{365}{mm}$ and $\alpha = 0.32$. The time $\Tilde{t}=t-t_{\text{impact}}$ is shown in each panel. Here $\Tilde{t}=0$ corresponds to the instant where the impact takes place, and thus (negative/positive) times correspond to (before/after) the impact takes places, respectively. The first four panels $\Tilde{t}\in[-50, -12.5] \text{ms}$ show the breaking process due to the interaction between the soliton and the beach. The next eight panels $\Tilde{t}\in[0, 87.5] \text{ms}$ correspond to the oscillation of the gas pocket, while for $\Tilde{t}>100 \ \text{ms}$, we observe that the gas pocket rises towards the free surface due to buoyancy. }
    \label{fig:wave_impact}
\end{figure*}

In \fref{fig:wave_impact}, we show a typical SIW experiment in the ATM. Here, a soliton is generated by the WM which propagates towards the so-called ``impact wall'' located at the other side of the flume. Due to the interaction with the beach the soliton undergoes breaking and the impact process begins. This is shown in the the first four panels of \fref{fig:wave_impact} for $\tilde{t}\in[-50,-12.5] \ \unit{\milli \second}$. Here $\tilde{t}=t-t_i$, where $t$ is time and $t_i$ is the time at which the impact occurs. Indeed, at $\tilde{t}=0$, the wave crest reaches the impact wall and a gas pocket is formed. As the impact process further develops, fluid is forced along the wall via two upwards and downwards jets (see for instance \fref{fig:wave_impact} at $\tilde{t}=\qty{12.5}{\milli \second}$). At the same time, the gas pocket experiences a series of compression and expansion cycles for $\tilde{t}>\qty{0}{\milli \second }$. This -- as will be shown later -- leads to an oscillatory response of the pressure inside the gas pocket, whose amplitude decays over time. Eventually, the oscillations fade out, buoyancy takes over around $\tilde{t}>\qty{100}{\milli \second}$, and the gas pocket rises towards the free surface.

\subsection{The pressure sensors of the ATM}
\label{sec:p_sensors}

The impact wall of the ATM is instrumented with one hundred \textit{Kistler type 601CAA} dynamic pressure sensors of 5.5 mm in diameter. \magenta{In particular, a single sensor measures the pressure in units of charge (\unit{\pico \coulomb}). This signal is then amplified and transformed to a voltage signal by a charge amplifier, after which the signal can be read by the measurement system. Every sensor has its own calibration value (in units of \unit{\pico \coulomb \per \bar}) which is provided by the manufacturer.} In our setup, one of these sensors is used as a trigger which leaves ninety-nine available sensors to measure the impact pressures. All sensors are flushed with the impact wall and measure the impact pressures as a function of time in units of bar and operate at a max. sampling rate of \qty{200}{\kilo\hertz}. In this work, the sampling rate is set to $\qty{100}{\kilo\hertz}$. \magenta{As will be shown later, the maximum measured frequency in this work is in the order of $\qty{150}{\hertz}$. When considering five oscillations, this sampling rate allow us to obtain $(5/\qty{150}{\hertz}) \times \qty{100}{\kilo \hertz} \approx 3333$ points which is sufficient to extract the dynamics of the oscillations.} The distribution of the sensors is such that each sensor is located at a unique height $z$. This feature allows for a better resolution of the impact pressures along the vertical direction -- as opposed to a single line of sensors that are stacked on top of each other. A sketch of the pressure sensor array can be seen in \fref{fig:pressure_sensors}a, where the 100 pressure sensors are depicted in an array that consists of 16 rows with 6 sensors in each row -- with an additional row of 4 sensors at the bottom. The vertical spacing for this array is $dz=\qty{0.83}{mm}$ (measured from center to center), while the horizontal spacing between sensors is $dy = \qty{10}{mm}$.

\subsection{Wave generation}
\label{sec:wave_generation}
The generation of the solitons is based on the works of \citet{guizien2002} and \cite{wu2016}. The goal here is to obtain the target (or steering) motion for the WM as a function of time $X_s\equiv  X_s(t)$. Essentially, the method is based on matching the paddle velocity of the WM at each position in time $t$ with the vertically averaged horizontal velocity of the wave $\tilde{u}$. This can be written as 

\begin{equation}
\frac{d X_s}{d t} = \Bar{u}(X_s,t).
\label{eq:wm_match}
\end{equation}

The solitary wave depth-averaged velocity $\Bar{u}$ in turn can be approximated by various theories. For instance, based on \citet{boussinesq1871} and \citet{rayleigh1876} formulations, \citet{goring1978} derived

\begin{equation}
    \Bar{u} = \frac{c \eta(\theta)}{h_\ell + \eta(\theta)},
\label{eq:avg_vel_soliton}
\end{equation}

\noindent where $\theta = ct - X_s$, $c$ is the wave speed, $\eta$ is the free surface wave elevation and $h_\ell$ is the water depth. Given $c$ and $\eta$ in \eref{eq:avg_vel_soliton}, the steering signal $X_s$ can be found by numerically integrating \eref{eq:wm_match}. In this study, both $c$ and $\eta$ are taken from the asymptotic solutions (up to third order) of \cite{grimshaw1971} (or \citet{fenton1972}) as follows

\begin{equation}
\begin{aligned}
k &= (3A/(4 h_\ell^3 ))^{1/2} \left(1 -\frac{5}{8}\alpha + \frac{71}{128} \alpha^2   \right), \\
c &= \sqrt{gh_\ell} \left( 1+ \dfrac{1}{2}\alpha - \frac{3}{20} \alpha^2 + \frac{3}{56} \alpha^3  \right),   \\
S &= \text{sech}(k(X_s- ct - x_0)), \\
T &= \tanh{(k(X_s- ct - x_0))}, \\
\eta &= h_\ell \left( \alpha S^2 - \frac{3}{4} S^2 T^2 \alpha^2 + (\frac{5}{8}S^2T^2-\frac{101}{80}S^4T^2)\alpha^3     \right),
\end{aligned}
\end{equation}

\noindent where $A$ is the soliton wave amplitude, $k$ is the wavenumber and

\begin{equation}
\alpha = \frac{A}{h_\ell},
\label{eq:alpha}
\end{equation}

\noindent is the parameter associated to weak nonlinearities \citep{wang2022}. Here, $x_0=-L_0=\qty{10}{m}$ is set to match the distance from the default zero position of the WM to the impact wall as suggested by \citet{wu2016}.  

With this formulation, the generation of $X_s$ is fully determined by only two parameters, namely $h_\ell$ and $\alpha$. These in turn determine the amplitude of the soliton $A$ via \eref{eq:alpha}. Note that, upon fixing either $\alpha$ or $h_\ell$, the wave generation reduces to a single control parameter. In order to show the influence of these two control parameters on the wave generation, it is useful to briefly study the GWS that can be generated by precisely fixing either $\alpha$ or $h_\ell$. We note that alternatively, one can choose $A$ and $h_\ell$ as the control parameters which in turn sets the value of $\alpha$. 

\begin{figure*}[ht]
    \centering
    \includegraphics[width = \linewidth]{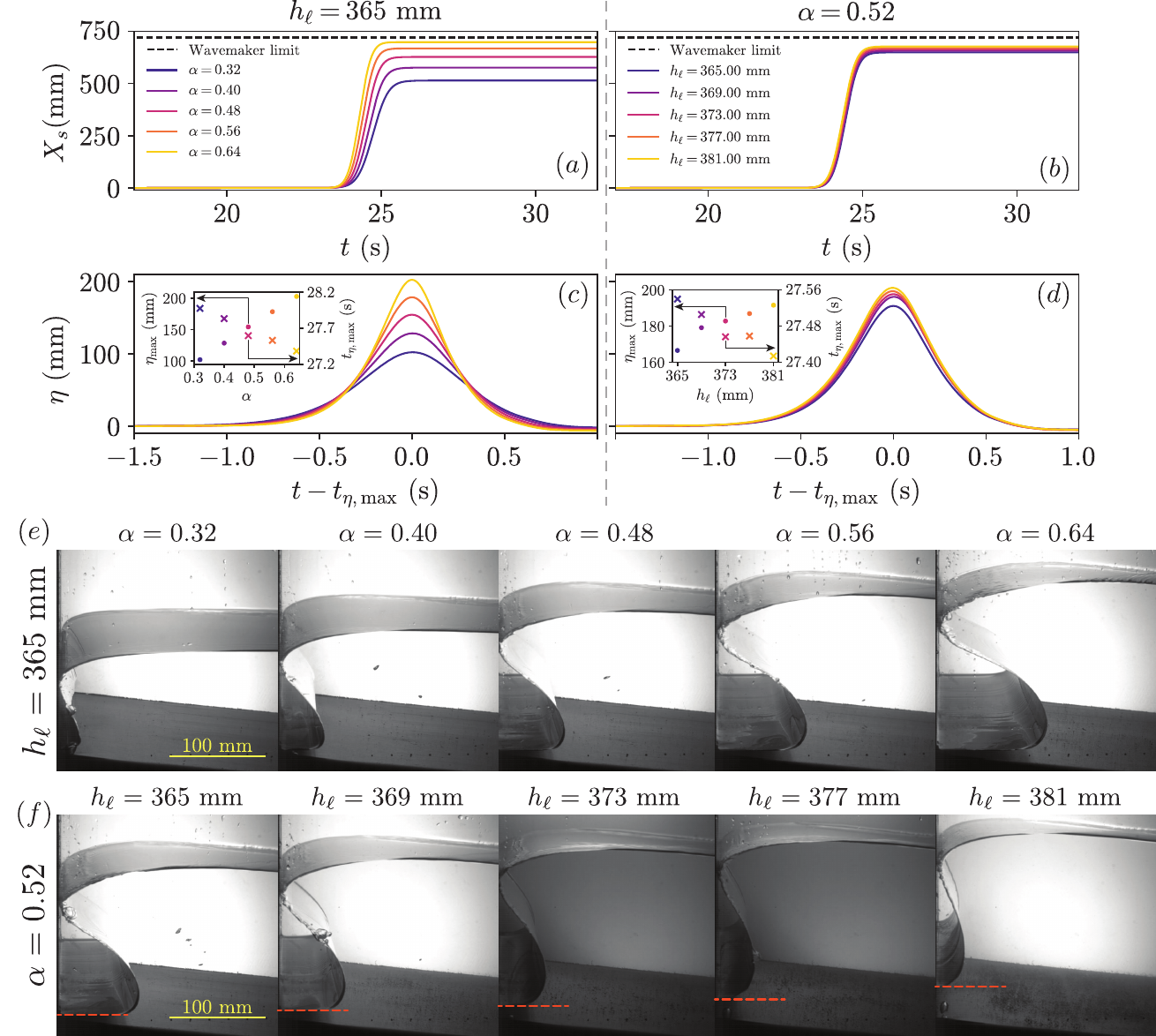}
    \caption{Wave generation in the ATM facility with solitons for various $h_\ell$ and $\alpha$. (a) Steering signal $X_s$ and (c) wave elevation $\eta$ measured at $x=\qty{2.76}{m}$ for fixed $h_\ell = 365 \ \text{mm}$ and varying $\alpha$.(b) Steering signal $X_s$ and (d) wave elevation $\eta$ measured at $x=\qty{2.76}{m}$ for fixed $\alpha = 0.52$ and varying $h_\ell$. (e) Breaking wave upon impact ($\tilde{t}=\qty{0}{\second}$) for the waves shown in (a,c) i.e. fixed $h_\ell$. (f) Breaking wave upon impact ($\tilde{t}=\qty{0}{\second}$) for the waves shown in (b,d) i.e. fixed $\alpha$. The red dashed line indicates the height of the wave trough upon impact. The lack of illumination in the third and fourth panels of (f) are due to a malfunction of the illumination system. The wave states in (e,f) are shown as magenta rectangles in \fref{fig:phase_space}, while the wave states in (b,d) are shown as purple rectangles in \fref{fig:phase_space}. The horizontal black dashed line in (a) and (b) corresponds to the max. stroke of the WM $X_{\max,\text{WM}}=\qty{720}{mm}$. A video of panels (e) and (f) figure can be found in the Supplementary Material as Video 1.}
    \label{fig:solitonsandhs}
\end{figure*}

Firstly, we show in \fref{fig:solitonsandhs}a, the calculation of the steering signal $X_s$ for fixed water depth $h_\ell = 365 \ \text{mm} $ and for various $\alpha$. The calculation reveals the typical ``signature'' associated to the generation of solitons – namely a sudden single stroke from rest up to a constant value which resembles a $\tanh$ function in time. Here, the signals are designed so the WM starts moving around $24$ s – this is done for triggering purposes of the measurement systems in the ATM. An increment in $\alpha$ translates into an increment in both the maximum amplitude of the stroke and the slope -- and thus the velocity at which the maximum stroke is reached as shown in \fref{fig:solitonsandhs}a for $t\in[24,25]$ s. The maximum stroke of the WM is limited by the mechanics and has a value of $X_{\max,\text{WM}}=\qty{720}{mm}$. This can be achieved by shifting the zero position of the WM to $\qty{-0.46}{\meter}$ as seen in \fref{fig:sketch_flume}a and thus, the effective distance that a soliton has to travel from the WM to the impact wall is $L_{\text{eff}}=L_0 + \qty{0.46}{\meter} = \qty{10.46}{m}$. The max. stroke of the WM imposes a first limit to the wave generation. Namely, that for a given $h_\ell$, the WM motion at its maximum amplitude should be less than $X_{\max,\text{WM}}$. 

To further illustrate the effect of fixed $h_\ell$ on the wave generation, we show in \fref{fig:solitonsandhs}c the corresponding wave elevation $\eta\equiv \eta(t)$ measured at $x=\qty{2.76}{m}$ with one of the ``low-speed'' cameras. Here, we also include snapshots of the GWS upon impact (\fref{fig:solitonsandhs}e) as obtained from the HS camera. Regarding the wave elevations, we observe that as $\alpha$ increases, the wavelet becomes narrower and the maximum wave amplitude $\eta_{\max}$ monotonically increases. In addition, we find that the propagation speed increases with increasing $\alpha$ (or $A$ in \eref{eq:alpha}) -- one of the most characteristic features of a solitary wave. The latter is evidenced by a decreasing $t_{\eta,\max}$, i.e. the time at which  $\eta_{\max}$ is measured. Upon inspection of the GWS in \fref{fig:solitonsandhs}e, we observe a dramatic influence of $\alpha$ on the GWS. Namely, that both the amplitude and the volume of entrapped gas monotonically increase with increasing $\alpha$. In addition, the wave crest becomes more narrow as $\alpha$ increases (smaller curvature). Lastly, we note that the wave trough appears not to move during the breaking process for all $\alpha$, i.e. no wave ``run-up'' is present.

Alternatively, we look now at the case of fixed $\alpha$ and varying $h_\ell$. In \fref{fig:solitonsandhs}b, we show firstly the corresponding steering signals $X_s$. In contrast to the case of fixed $\alpha$, here we observe that $X_s$ remain relatively unchanged within this small range of $h_\ell$. Nonetheless, the wave elevations shown in \fref{fig:solitonsandhs}d, also reveal that both $\eta_{\max}$ and the propagation speed increase with $h_\ell$ -- although with a smaller variation. Additionally, we observe that the wavelets slightly broaden with increasing $h_{\ell}$. When looking at the snapshots upon impact in \fref{fig:solitonsandhs}f however, we clearly observe again a strong effect on the GWS. Here, the wave amplitude slightly increases (similarly to the wave elevation) and the volume of entrapped gas monotonically decreases. Additionally, we find that the wave crest broadens with increasing $h_\ell$ (larger curvature) and that here, a  wave ``run-up'' can be observed for $h_\ell>373 \ \text{mm}$. We note that these trends (fixed $\alpha$) are essentially the opposite as the ones for fixed $h_\ell$ and will be further discussed in the next section. In Video 1 of the Supplementary Material, we show the wave impacts shown in \fref{fig:solitonsandhs}e,f  for $\tilde{t}=[-50, 175] \ \text{ms}$.

\subsection{The phase space of wave generation}
\label{sec:phase_space}
In the previous section, we showed that fixing either $\alpha$ or $h_\ell$ has a dramatic influence on the GWS. In this section, we further explore this feature by placing these observations in the context of a two-dimensional region defined precisely by both $\alpha$ and $h_\ell$ and study their dependency therein. In the context of fluid physics, such exercise is often referred to as ``exploring the phase space". In this work, we adopt this methodology and define ``wave states" as pairs of ($h_\ell,\alpha$) within this region. The advantage of using this representation is twofold: i) Firstly, it allows us to gain a better understanding on the kind of GWSs (and thus gas pockets) that we can generate. ii) Secondly, it allows us to properly define a dataset that can be used to train the ML model. The procedure to construct such a phase space is described as follows.

\begin{figure}[ht]
    \centering
    \includegraphics[width = \linewidth]{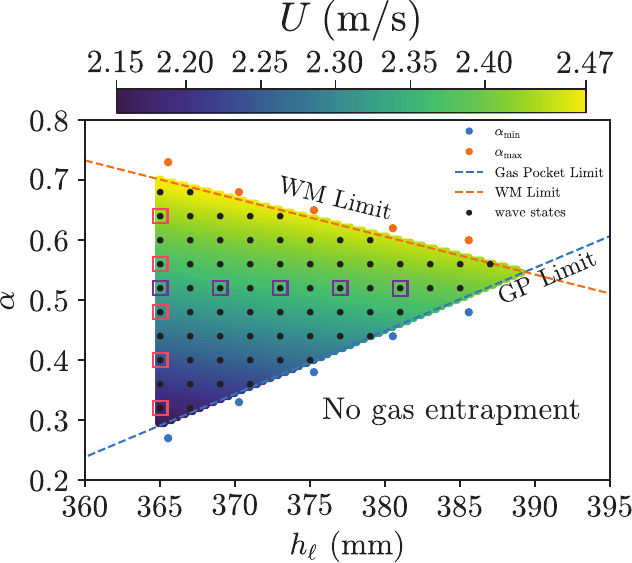}
    \caption{Phase space of wave generation. Here, wave states inside the triangular region bounded by $h_\ell=\qty{365}{mm}$, the WM and the GP limits lead to breaking waves that are able to entrap gas. The blue/orange solid points represent $\alpha_{\min}/\alpha_{\max}$, respectively. The orange/blue dashed lines are the WM limit and the GP limit, respectively. The black solid points are the wave states where the experiments are performed. The magenta rectangles (fixed $h_\ell=\qty{365}{mm}$) corresponds to (a)(c)(e) shown in \fref{fig:solitonsandhs}, while the purple rectangles (fixed $\alpha=0.52$) corresponds to (b)(d)(f) shown in \fref{fig:solitonsandhs}. The colorbar represents the magnitude of the propagation speed of the soliton $U=\sqrt{g h_\ell (1+\alpha)}$. The min. and max. available propagation speeds are shown as the min. and max. in the colorbar. A video of the different GWS that can be obtained for every wave state can be found in the Supplementary Material as Video 2.}
    \label{fig:phase_space}
\end{figure}

We start by noting that similar experiments with solitons by \citet{ezeta2023} at fixed $h_\ell=365 \ \text{mm}$ reveal that when $\alpha$ is sufficiently small ($\alpha\approx 0.30$ and below), no breaking occurs and thus no gas pocket formation is observed. Rather, the GWS that is obtained in this case corresponds to a ``jet'' that is forced upwards along the wall. Only when $\alpha$ is increased beyond a certain value, wave breaking occurs and gas entrapment is observed. This occurs for $\alpha \geq \alpha_{\min}$, where $\alpha_{\min}$ is precisely the smallest $\alpha$ for which gas entrapment is observed. In addition, for a given intermediate value ($0.3<\alpha<\alpha_{\min}$) one can even obtain a ``flip-through'' type wave. Note that the precise value of $\alpha$ at which these transition occurs is  $h_\ell$-dependent. This control of the GWS by manipulating $\alpha$ (for fixed $h_\ell$) is analogous to manipulating the value of the focusing point in the context of generating a focusing wave as shown by \citet{hofland2011}. In this spirit, for a given $h_\ell$, we manually find $\alpha_{\min}$ for various $h_\ell$. These wave states are highlighted in solid blue points in \fref{fig:phase_space}, where we find that $\alpha_{\min}$ monotonically increases with increasing $h_\ell$. Similarly, for the same values of $h_\ell$, we now look at the maximum value of the wave steepness $\alpha_{\max}$ that can be executed by the WM for a given $h_\ell$. As described in \sref{sec:wave_generation}, this value is limited by $X_{\max, \text{WM}}$ for a given $h_\ell$ – these states are highlighted in solid orange points in \fref{fig:phase_space}. Here, we find that $\alpha_{\max}$ monotonically decreases with increasing $h_\ell$.

Next, we linearly interpolate both $\alpha_{\min}(h_\ell)$ and $\alpha_{\max}(h_\ell)$ and add/subtract an arbitrary value of $c=\qty{0.02}{mm}$ to both fits, respectively – these fits are shown with the orange and blue dashed lines in \fref{fig:phase_space} and are called WM Limit and GP limit, respectively. We note that while both $\alpha_{\min}$ and $\alpha_{\max}$ are likely not linear functions of $h_\ell$, the usage of these linear limits allow us to introduce a rather simple ``safety factor''  that guarantees that above the GP limit, wave breaking will occur and thus gas will be entrapped. In this way, we define a ``phase space'' of wave generation as the region bounded by the minimum water depth used in this work, i.e. $h_\ell =\qty{365}{mm}$, and both the GP and WM (i.e. shaded triangle in \fref{fig:phase_space}). Thus, wave states that lie inside of this triangular region are guaranteed to result in wave breaking and therefore a gas pocket.

In addition, it is also insightful to investigate the propagation speed of the soliton in phase space. We estimate this speed as $U=\sqrt{g (A + h_\ell)}=\sqrt{gh_\ell (1 + \alpha)}$, where $g$ is the gravitational acceleration and where we have used \eref{eq:alpha} to write $U$ in terms of $h_\ell$ and $\alpha$ \citep{boussinesq1871,rayleigh1876,serre1953,goring1978}. The propagation speed of the soliton $U$ is also shown in \fref{fig:phase_space}, where we see the formation of a gradient of $U$ towards larger values of both $h_\ell$ and $\alpha$. In phase space, the minimum and maximum values of $U$ are located at the lowest water depth $h_\ell = \qty{365}{mm}$. These values are $\qty{2.15}{m/s}$ for $\alpha=0.295$ and $\qty{2.47}{m/s}$ for $\alpha=0.7$. We note that while in principle two different wave states could potentially lead to very similar GWS (see for instance the third panel \fref{fig:solitonsandhs}(c) and the second panel in \fref{fig:solitonsandhs}(d)), the fact that the wave kinematics is different (different $U$) for those two wave states, guarantees that every wave state in phase space is unique. This is relevant as we do not want to introduce biases due to the GWS when training the ML model. \magenta{For completeness, we note that a Mach number in the gas phase can be estimated by using $M_g = U / c_g $. As $c_g\approx \qty
{343}{\meter \per \second}$ remains effectively constant for all experiments and the difference between the max. and min. propagation speed is small, then the variation of Mach number is in the order of $\Delta U /c_g = (\qty{2.47}{\meter \per \second} - \qty{2.15}{\meter \per \second}) / \qty{343}{\meter \per \second} \approx 9.32 \times 10^{-4}$. Due to this small variation, we do not investigate the role of gas compressibility in this work. }

Finally, we construct the dataset by selecting 67 wave states within phase space. This is done with a resolution for the water depth and wave steepness of $d h_\ell=2 \ \text{mm}$ and $d \alpha = 0.04$, respectively, which we show in \fref{fig:phase_space} as black solid points. The minimum $\alpha$ and $h_\ell$ used in this work is $0.32$ and $h_\ell=\qty{365}{\milli \meter}$, respectively. In Video 2 of the Supplementary Material we show the different GWS that can be obtained for every wave state explored in phase space.

\subsection{Quantifying the gas pocked dynamics: the target outputs}
\label{sec:outputs}    

In order to quantitatively describe the dynamics of the gas pocket oscillation for every wave state, we have selected six output variables which are described in this section. 

\begin{figure*}[ht]
    \centering
    \includegraphics[width = \linewidth]{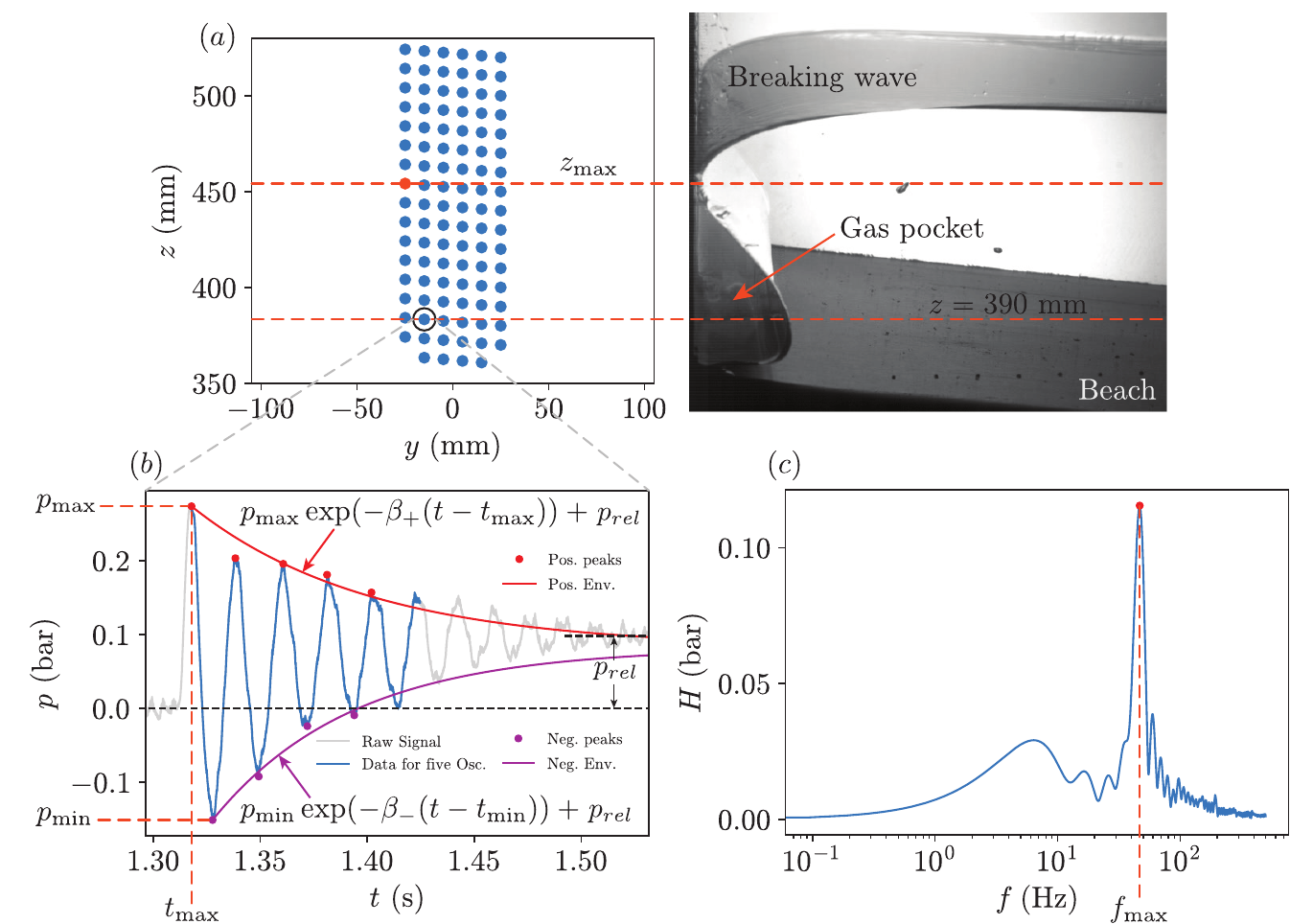}
    \caption{Summary of the output variables that quantify the dynamics of the gas pocket oscillation. (a) Sketch of the pressure sensor array located at the impact wall. These array consists of 100 pressure sensors which are vertically spaced by $dz=\qty{0.83}{mm}$ and $dy=\qty{10}{mm}$. The values of $z$ are measured with respect to the floor of the flume. This panel includes a snapshot of a breaking wave where the red dashed-line represents the height $z_{\max}$, i.e. the height at which the maximum pressure $p_{\max}$ is recorded and thus where the ELP1 is most likely to occur. (b) Pressure signal as a function of time for $z=\qty{390}{mm}$ (16th sensor). Here, the data in gray represents the raw data as obtained from the sensor, the data in blue is the same data albeit in the interval $t\in[t_{\max},t_{\max}+5/f_{\max}]$. Within this time interval, the positive ($t_+,p_+$) and negative peaks ($t_-,p_-$) used for the calculation of the envelopes are shown as red and purple points, respectively. The positive and negative envelopes are shown as red and purple solid lines, respectively. (c) Zero padded Fast Fourier Transform of the data shown in blue in (b). The dominant frequency is shown as $f_{\max}$. }
    \label{fig:pressure_sensors}
\end{figure*}

Prior to impact, the wave crest velocity becomes larger than the velocity of the free surface close to the wave trough which leads to the impact on the solid wall. In the context of Elementary Loading Processes (ELPs) \citep{lafeber2012}, this first point of contact with the solid wall is typically associated to the so-called ELP1 (i.e. ``direct impact'') and is characterized by a large amplitude, short rise time pressure signal. In our experiments, the ELP1 is approximately captured by the pressure sensor that measures the largest pressure. This is located at the height $z_{\max}$ which lies very close the point of maximum curvature in the wave crest as can be seen in \fref{fig:pressure_sensors}a. As time evolves, the pressure sensors which are below the location of the ELP1 ($z<z_{\max}$) start to capture both the contribution of the jet traveling downwards and the gas pocket. Similarly, the pressure sensors that lie above $z>z_{\max}$ mostly measure the loading of the jet running upwards and are thus ignored in this work. We note that in the context of ELPs, the loading of these traveling jets define the so-called ELP2 and that the oscillating pressure at the wall due to the gas pocket is associated to the so-called ELP3. 

A typical pressure signal associated to the gas pocket oscillation (or ELP3) can be seen in \fref{fig:pressure_sensors}b. This measurement corresponds to the wave state ($h_\ell=\qty{365}{mm}$, $\alpha=0.40$) for the 16th sensor ($z=\qty{390} \ \text{mm}$) which is also highlighted with a black circle in \fref{fig:pressure_sensors}a. Here, the height $z$ is measured with respect to the floor of the flume. The pressure signal reveals a maximum value of $p_{max}$ measured at $t_{\max}$ that can be associated to the maximum compression phase of the gas pocket. Similarly, a minimum value of the pressure $p_{\min}$ is measured for some $t>t_{\max}$ which can be attributed to the maximum expansion of the gas pocket. We note that $t_{\max}$ is measured with respect to an arbitrary triggering time $t_{p}=\qty{30.1}{s}$. In other words, the measurement of the impact pressures starts $t_p$ seconds after the WM signal is executed which occurs at $t\approx \qty{1}{\second}$ in \frefs{fig:solitonsandhs}a,b. In principle, the time scale $t_{\max}$ contains both information from the propagation speed (the time it takes for the soliton to reach the impact wall) and the rise time, which is often defined as $t_r=2(t_{\max} - t_{\max/2})$, where $t_{\max/2}$ is the time at which half of the maximum pressure is measured. However, as $t_{max} \gg t_r$, we expect -- as will be shown later -- that the propagation speed is mostly responsible for the trends of $t_{\max}$ in phase space. Additionally, we extract the dominant frequency of oscillation $f_{\max}$ associated to this motion, which can be obtained via Fourier analysis. In particular, we use the zero-padded Fast Fourier Transform of the pressure signal signal $H$ for the first five oscillations of the gas pocket ($t\in[t_{\max},t_{\max} + 5/ f_{\max}]$), which is shown as the blue solid line in \fref{fig:pressure_sensors}a. For every wave state, we choose only five periods to estimate the frequency and the decay rates, as the HS recordings reveal that after this time, the gas pocket moves upwards due to buoyancy (see last four panels of \fref{fig:wave_impact}). Furthermore, we quantify the decay rates (positive and negative) within the same time interval by using a similar approach as \cite{bogaert2018}, where an exponential decay fit is used. This methodology is as follows: (i) First, we subtract the relaxation value $p_{rel}$ from the pressure signal such that the oscillations are centered around zero – this value can also be seen in \fref{fig:pressure_sensors}b. (ii) Next, we find the local minima $(t_{-},p_{-})$ and maxima $(t_{+},p_{+})$ of the pressure signal in the interval $t\in[t_{\max},t_{\max} + 5/ f_{\max}]$ -- these are labeled as ``Neg. peaks'' and ``Pos. peaks'' in \fref{fig:pressure_sensors}b, respectively.  (iii) Next, we fit the following functions to $(t_+,p_{+})$ and $(t_-,|p_{-}|)$

\begin{eqnarray}
P_{+}(t) &=& P_{\max} \exp(-\beta_+(t-t_{\max})) \\
P_{-}(t) &=& -|P_{\min}| \exp(-\beta_-(t-t_{\min})),
\label{eq:pos_fit} 
\end{eqnarray}

\noindent where $P_{-}$, $P_{+}$ are the negative and positive envelopes, respectively, and $\beta_{-}$, $\beta_{+}$ are the negative and positive decay rates, respectively. This formulation guarantees that as $t\to \infty$, the pressure is zero (as we have subtracted the relaxation value $p_{rel}$) and that $P_{+}(t_{\max}) = p_{\max}$ and $P_{-}(t_{\min}) = p_{\min}$. (iv) Finally, we add back the relaxation factor $p_{rel}$ to both $P_{-}$ and $P_{+}$. In \fref{fig:pressure_sensors}b, the negative envelope $P_{-}$ is shown as a purple line, while $P_{+}$ is shown with a red solid line. 

From the former analysis, we obtain finally a list of six scalars -- a summary of which can be seen in \tref{tab:outputs}. \magenta{As essentially every pressure sensor leads to six scalars, we perform an additional average that leads to only six scalars per experiment. We do this by introducing a ``gas pocket average'' $\langle \rangle_{gp}$. This average consists of an axial average (i.e. over the height $z$) of 10 sensors that lie within the gas pocket. As the GWS upon impact depends on the wave state, the sensors that define $\langle \rangle_{gp}$ must be chosen with care. This is further explained in \sref{sec:outputs_phase_space}. The gas pocket average of these scalars is precisely what is predicted by the ML models in this work.}

\begin{table}[h]
    \centering
    \resizebox{\linewidth}{!}{%
    \begin{tabular}{l|l|l}

         Output & Units & Description  \\
         \hline
         $p_{\max}$& \unit{bar} & Maximum pressure.\\
         \hline
         $p_{\min}$& \unit{bar} & Minimum pressure \\
         \hline
         $t_{\max}$& \unit{\second} & The time at which the max. pressure is measured\\
         \hline
         $f_{\max}$& \unit{\hertz} & The dominant frequency of oscillation. \\
         \hline
         $\beta_+ $& \unit{\hertz} & The positive decay rate. \\   
         \hline
         $\beta_- $& \unit{\hertz} & The negative decay rate. \\
        
    \end{tabular}
    }
    \caption{The list of target outputs that define the dynamics of the gas pocket in this work, see \fref{fig:pressure_sensors} for a graphical description.  }
    \label{tab:outputs}
\end{table}

\subsection{Experimental procedure}
The procedure to execute a typical experiment is as follows. For a given wave state, the corresponding steering signal $X_s$ is loaded onto the WM and the measurement systems are simultaneously initiated. These systems comprise of the WM encoder, LS cameras, HS cameras and impact pressures. The duration of every experiment is approximately 70 s, after which the wavemaker is moved back to its zero position $x_0=\qty{-0.46}{m}$. A waiting time of 10 min is introduced in between experiments to allow the water surface to go back to equilibrium. The training set consists of 67 wave states as shown in \fref{fig:phase_space}. For every wave state, we perform three repetitions which yields a total of 201 experiments. Finally, for every experiment we collect the following datasets:

\begin{enumerate}
    \item Wavemaker motion as a function of time $X\equiv X(t)$  as measured by the encoder located at the WM.
    \item Wave elevations $\eta$ at three different positions along the flume as measured by the low-speed cameras
    \item High-speed recording of the wave impact near the impact wall
    \item Impact pressures measured by 99 pressure sensors as described in \sref{sec:p_sensors}
    \item Measurement of the actual water depth $h_{\ell}$ and liquid temperature $T_\ell$.    
\end{enumerate}

\section{Material and Methods: Machine Learning Regression}
\label{chap:materials_ml}

In this section we describe the problem set-up to predict the target outputs defined in \sref{sec:outputs} using the ML techniques. As mentioned earlier, the focus of the present work is on the cLSTM model. However, we also train a MLP model with the same data and use it as a baseline. Therefore, as the input to the regression, we consider the following two perspectives:

\begin{enumerate}
    \item MLP -- The parameter $\alpha$ and the water depth $h_\ell$ are considered as input. In this case, the MLP model is essentially learning the surfaces shown in \fref{fig:exp}.  
    \item cLSTM -- A time series comprised of two snapshots of the breaking waves before impact is considered as input. The snapshots are obtained with the high-speed camera. Here, we use the convolutional long short-term memory (ConvLSTM) structure of \citet{shi2015convolutional} as the basis for the time series regression model.
\end{enumerate}

\magenta{The hyperparameters of both models are optimized using a grid-search technique with 5-fold cross-validation.}
Both approaches have different data preparation procedures, model architectures and hyperparameter sets. 
%As such, we describe them in separate sections. 
In this section we describe only the details of the cLSTM model. For the sake of conciseness, the details of the MLP model are given in \sref{app:theMLPmodel}. 

\begin{figure*}[ht]
    \centering
    \includegraphics[width=\textwidth]{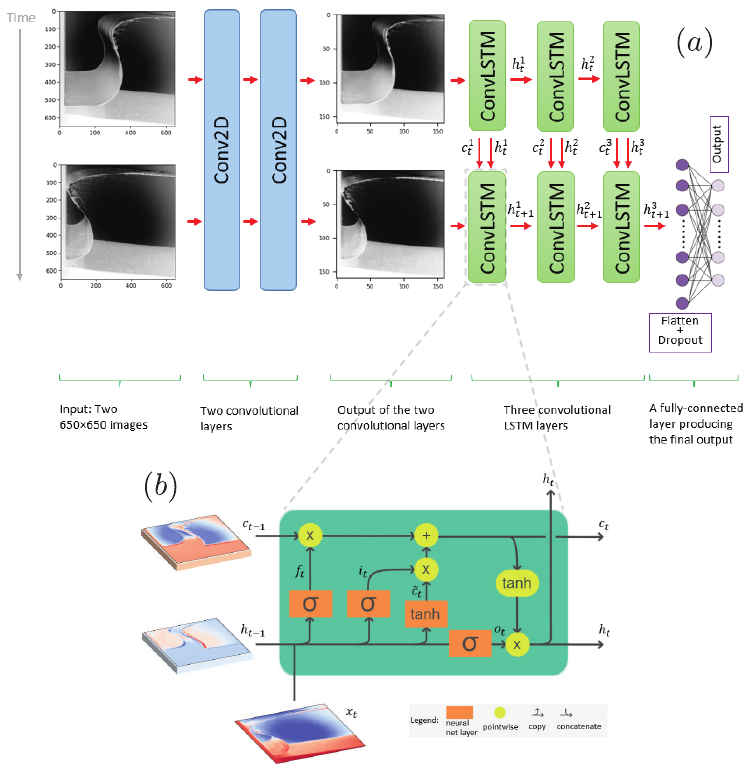}
    \caption{(a) Architecture of the proposed cLSTM model. The superscript in the outputs of the ConvLSTM layers indicates the layer number, and subscript indicates time. (b) \magenta{Architecture of a ConvLSTM cell.}}
    \label{fig:convlstmDiag}
\end{figure*}

In \fref{fig:convlstmDiag}, we show the architecture of the cLSTM model. Our model takes as input two snapshots of a breaking wave before impact. Each snapshot has a resolution of $650 \times 650$ pixels. For each wave state, the first snapshot is taken at impact and the second one is taken 140 frames earlier which corresponds to $\approx \qty{35}{\milli \second}$. \magenta{This choice is based on a preliminary study, where the same model is trained separately for different number of frames and time intervals. Ultimately, the choice of two snapshots separated by $\qty{35}{\milli \second}$ yields the best prediction accuracy and is thus selected.} The snapshots are in gray-scale (one channel only), and their pixel values are scaled to the range of $[-1, 1]$. Therefore, the input to the neural network has the shape of $2 \times 1 \times 650 \times 650$, where $2$ indicates the length of the time series, and $1$ indicates that there is only one channel in the snapshots.

Each snapshot is first put through two 2D convolutional layers, each with a kernel of $5 \times 5$ and a stride of $2 \times 2$. The first layer has 8 filters and the second layer 1 filter. The Sigmoid activation function is used after each layer. Hence, the output of the convolutional layers has the shape of $2 \times 1 \times 160 \times 160$. The purpose of the two convolutional layers is to reduce the image size while retaining as much contextual information as possible. The two convolutional layers are followed by three ConvLSTM layers, which combines the temporal and spatial features of the data to achieve simultaneous extraction of spatiotemporal features. In each ConvLSTM cell illustrated in \fref{fig:convlstmDiag}b, the following calculations take place 

\begin{equation}
\begin{aligned}
    {i_t} & =\sigma \left( {{W_{xi}} * {x_t} + {W_{hi}} * {h_{t - 1}} + {b_i}} \right),\\
    {f_t} & =\sigma \left( {{W_{xf}} * {x_t} + {W_{hf}} * {h_{t - 1}} + {b_f}} \right),\\
    {o_t} & =\sigma \left( {{W_{xo}} * {x_t} + {W_{ho}} * {h_{t - 1}} + {b_o}} \right),\\   
    {{\tilde c}_t} & = \tanh \left( {{W_{xc}} * {x_t} + {W_{hc}} * {h_{t - 1}} + {b_c}} \right),\\    
    {c_t} & ={f_t} \circ {c_{t - 1}} + {i_t} \circ {{\tilde c}_t},\\ 
    {h_t} & ={o_t} \circ \tanh \left( {{c_t}} \right), 
\end{aligned}
\label{eq:convlstm}
\end{equation}

\noindent where ${c_{t - 1}}$ and ${h_{t - 1}}$ are the previous cell and hidden state, respectively, $x_t$ is the time-series data at time $t$, and $*$ and $\circ$ denote the convolution operator and the Hadamard product (element-wise product), respectively. The cell takes these three inputs (${c_{t - 1}}$, ${h_{t - 1}}$, $x_t$) and produce the new cell state ${c_t}$ and hidden state ${h_t}$ as its outputs. The key to LSTM's design is the cell state running at the top of the diagram which carries information from one cell to the next. It is possible to remove or add information to the cell state, which is regulated by structures called ``gates''. There are three gates controlling the flow of information in a cell; ${i_t}$ is the input gate, ${f_t}$ is the forget gate, and ${o_t}$ is the output gate. Each gate has a sigmoid activation which outputs numbers between zero and one, describing how much information should be let through. A value of zero means ``let nothing through'' while a value of one means ``let everything through''. $W$ indicates the convolution kernel, whose subscript indicates what it convolves for which output. For example, ${{W_{xi}}}$ suggests that the convolution operation is applied to $x$ for the calculation of the i-th input gate. Observing \eref{eq:convlstm}, we can see that the convolution operation is applied both at the input-to-state transitions and at the state-to-state transitions. In this way, the future state of a certain location in the grid is determined by the inputs and past states of its local neighbors. In calculating the new cell state ${c_t}$, the forget gate $f_t$ determines what information will be discarded from the previous cell state ${c_{t - 1}}$. The next step is to obtain the new information to store in the cell state. Here, a $\tanh$ layer creates new candidate values ${{\tilde c}_t}$, that could be added to the state. This is then multiplied by the input gate $i_t$ which determines how much to update each state value. Finally, the output $h_t$ is calculated based on the new cell state $c_t$ multiplied by the output gate $o_t$. The output gate determines how much of each value in the new cell state will be outputted using its sigmoid activation. The $\tanh$ activation working on the cell state pushes the values between -1 and 1. Each ConvLSTM layer has a hidden state dimension of 16. The input-to-state and state-to-state convolutional layers have the same kernel size of $7 \times 7$, and strides of $2 \times 2$ and $1 \times 1$, respectively. The input to both convolutional layers are first sent through a dropout layer $(p=0.3)$.

Since we aim to regress towards an array of six values (see \tref{tab:outputs}), a fully-connected layer is added as the output layer of the neural network. The output hidden state of the third ConvLSTM layer $(h_{t + 1}^3)$ is first flattened, then sent through a dropout layer $(p=0.3)$, and then through the fully-connected layer which finally outputs the six scalars. We refer to the complete model as cLSTM with 280K trainable parameters. The cLSTM is trained for 3500 epochs with a learning rate of $0.0003$ using the ADAM optimizer.

\section{Results: Experiments}
\label{chap:results_exp}

In this section, we describe the major findings regarding the experiments. In particular, we focus on the dependency of the six output variables that describe the gas pocket dynamics and explain quantitatively their dependency in phase space.

\subsection{The outputs in phase space}
\label{sec:outputs_phase_space}

\begin{figure*}[ht]
    \centering
    \includegraphics[width = \linewidth]{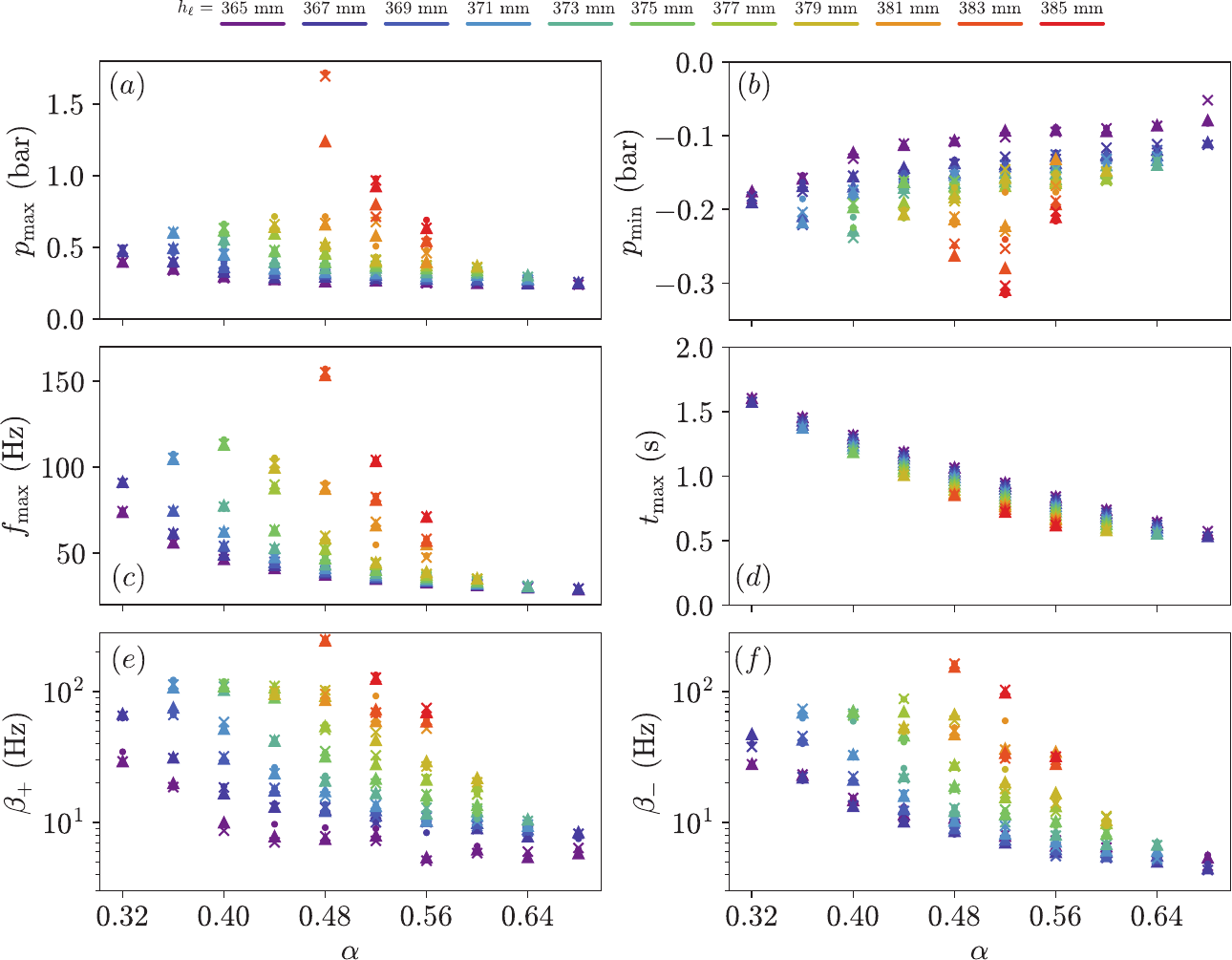}
    \caption{Gas pocket average $\langle \rangle_{gp}$ of the target outputs as a function of $\alpha$ and  $h_\ell$ for all three repetitions. (a) $p_{\max}$, (b) $p_{\min}$, (c) $f_{\max}$, (d) $t_{\max}$, (e)  $\beta_{+}$, and (f) $\beta_{-}$. The colors represent the water depth $h_\ell$ as described by the legend. The different markers for a given $\alpha$ and $h_\ell$ represent repetitions of that wave condition. For a graphical overview of how these quantities are obtained, we refer the reader to \fref{fig:pressure_sensors}. }
    \label{fig:avg_output}
\end{figure*}

\magenta{As mentioned at the end of \sref{sec:outputs}, the vertical region along the wall that defines the gas pocket average $\langle \rangle_{gp}$ is wave-state dependent. Thus, for every wave state, we look at the profiles of the maximum pressure $p_{\max}(z)$ and look for plateau regions in $z$ where $p_{\max}$ is uniform (i.e independent of $z$). Once this region in $z$ is known, we take 10 sensors within and perform the average $\langle \rangle_{gp}$ there.} In favor of maintaining a clearer notation, we have opted to omit the operation $\langle \rangle_{gp}$ throughout the rest of the manuscript. In \fref{fig:avg_output}, we show the gas pocket average $\langle \rangle_{gp}$ of all target outputs for all explored wave states and repetitions.

Firstly, \magenta{we discuss the gas pocket average of the max. pressure as a function of $\alpha$ in \fref{fig:avg_output}a}. When $h_{\ell}$ is fixed, we observe a monotonic decrease of $p_{\max}$ for increasing $\alpha$ -- this observation is obtained for all values of $h_{\ell}$. Conversely, when $\alpha$ is fixed, a monotonic increase of $p_{\max}$ with $h_\ell$ is revealed -- similarly, this is observed for all values of $\alpha$. Note that similar trends are obtained from the gas pocket average of the min. pressure $p_{\min}$ shown in \fref{fig:avg_output}b. In addition, note that the measurements are fairly repeatable -- except for $p_{\max}$ of the wave state ($\alpha=0.48$, $h_\ell = \qty{383}{\milli \meter}$). Interestingly, this wave state lies very close to the GP limit and has a very small gas pocket. This can be explained by the fact that  wave states that are closer to the GP limit, resemble more flip-through impacts where the variability of the impact pressure is large \citep{lugni2006,hofland2011}. Next, in \fref{fig:avg_output}c, we show the gas pocket average of the dominant frequency $f_{\max}$, where a similar trend as for the pressures is observed. Namely, that $f_{\max}$ monotonically decreases with increasing $\alpha$ for all $h_\ell$ and that $f_{\max}$ increases with increasing $h_\ell$ for all $\alpha$. 

In \fref{fig:avg_output}d, we show the gas pocket average of $t_{\max}$, where we also observe a monotonic decrease with increasing $\alpha$. This is explained by noting that for a soliton, an increment in $\alpha$ (for fixed $h_\ell$) translates into a monotonic increment of the amplitude according to \eref{eq:alpha}, and thus the propagation speed (see \frefs{fig:solitonsandhs}a,c and \fref{fig:phase_space}). In turn, an increment in the propagation speed yields a shorter ``arrival time'', i.e. the time it takes for the soliton to propagate along the extent of the flume. Indeed, $t_{\max}$ is a good indicator of the arrival time, as all experiments are triggered with the same internal clock. A similar argument can be used to explain the case of fixed $\alpha$, where we see that $t_{\max}$ decreases for increasing $h_\ell$. Here, the soliton propagates faster due to the increment in $h_\ell$. This figure reveals yet again the consistency of the wave generation behind the experiments in this work.

Finally, in \frefs{fig:avg_output}e,f, we show the average positive and negative decay rates, respectively. Here, we observe once more a decrease for both $\beta_+$ and $\beta_{-}$ for increasing $\alpha$ and for all $h_\ell$. This behavior can be explained by noting that -- as discussed previously -- both the amplitude (in absolute value) and the frequency are decreasing functions of $\alpha$ for a fixed $h_\ell$. In turn, it follows that the decay rates must decrease as the decay rates effectively quantify the magnitude of the local slope associated to either envelope.  As an illustration of this, we refer the reader to \fref{fig:pressure_sensors}b. On the other hand, we observe that when $\alpha$ is fixed, both decay rates increase for all water depths $h_\ell$ -- which can be explained by the same geometrical argument. Furthermore, we observe that the values of $\beta_+$ are larger than those of $\beta_-$. This can be attributed to the fact that both $|p_{\max}|>|p_{\min}|$ for all wave states, and since both quantities are included in the fitting of the exponential functions, we expect thus that $\beta_+$ decays faster than $\beta_-$.

\begin{figure*}[t!]
    \centering
    \includegraphics[width = \linewidth]{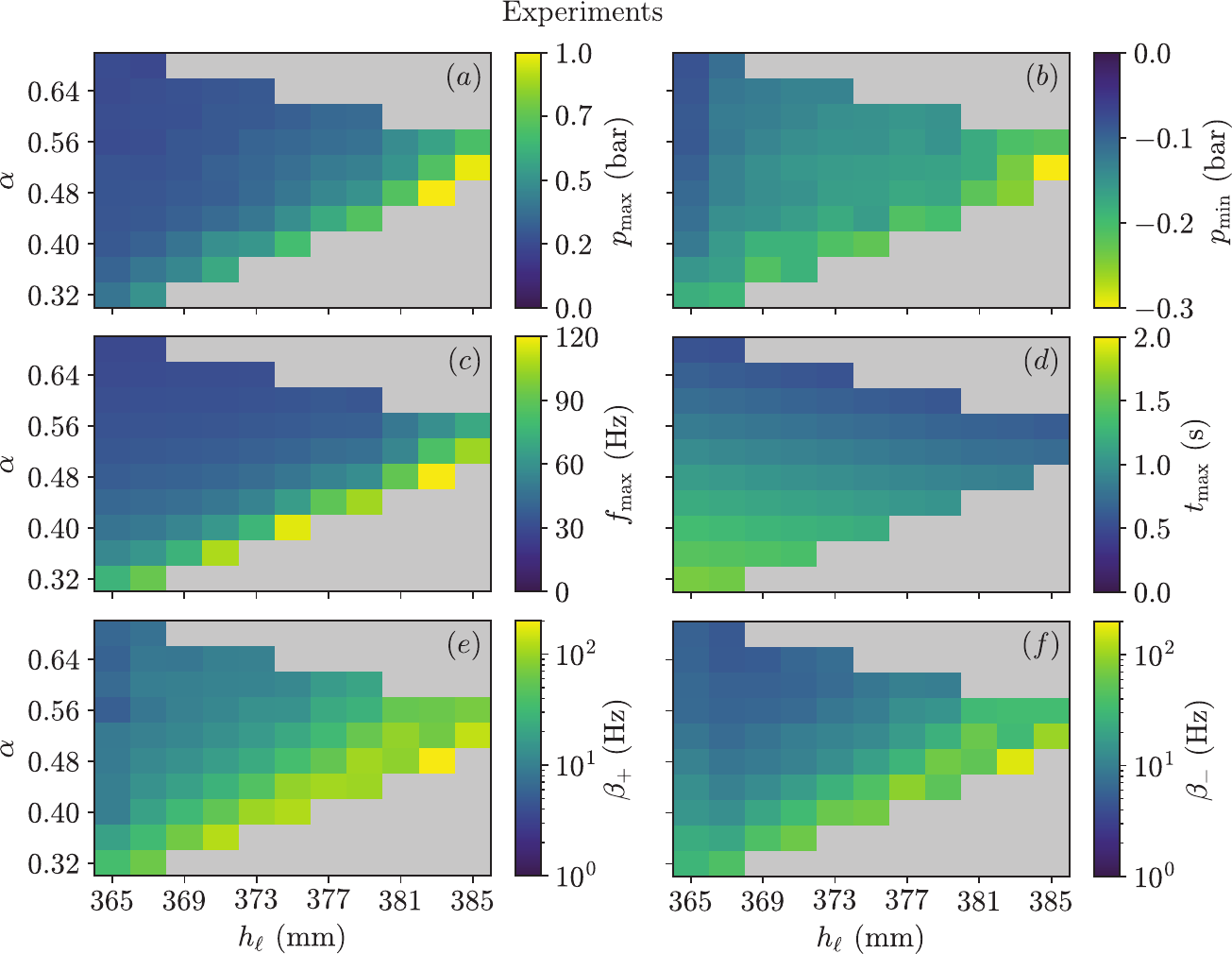}
    \caption{Gas pocket average $\langle \rangle_{gp}$ of the target outputs in phase space. The data shown here corresponds to the experiments. (a) $p_{\max}$, (b) $p_{\min}$, (c) $f_{\max}$, (d) $t_{\max}$, (e) $\beta_{+}$, and (f) $\beta_{-}$. In every panel, the color represents the magnitude of that quantity as described by the corresponding colorbar.}
    \label{fig:exp}
\end{figure*}

In order to summarize the trends shown in this section, we show in \fref{fig:exp} an alternative representation of the averaged outputs shown in \fref{fig:avg_output}. Here, we show all target outputs in phase space, i.e. as a function of both $\alpha$ and $h_{\ell}$ although only for the first repetition. Note that in this representation, one recovers the trends that were previously discussed for all outputs by either fixing $\alpha$ or $h_\ell$. As a consequence of this, one can clearly observe that the quantities  develop very well-defined gradients. In particular, the gradients of $p_{\max}$, $p_{\min}$, $f_{\max}$, $\beta_{+}$ and $\beta_-$ are directed towards smaller values of $\alpha$ and larger values of $h_\ell$. And these in turn, reach their maximum values at the GP limit for a given $h_\ell$. We note that all wave states that define the GP limit are characterized by small gas pockets as outlined in \sref{sec:phase_space}. Thus, our experiments are consistent with the general notion in the literature that small gas pockets generate larger impact pressures than those generated by large gas pockets \citep{hofland2011,bredmose2015}. 

In summary, we find that in phase space, the gradient of $t_{\max}$ is a direct consequence of the gradient of $U$. Similarly the gradients of $\beta_+$ and $\beta_-$ can be attributed to the combined behavior of the gradients of $p_{\max}$, $p_{\min}$ and $f_{\max}$. The mechanisms responsible for these last three gradients however, have not yet been discussed. In the following section, we focus on them and provide a qualitative description that can explain the reported trends.

\begin{figure}[ht!]
\centering
\includegraphics[width = 0.95\linewidth]{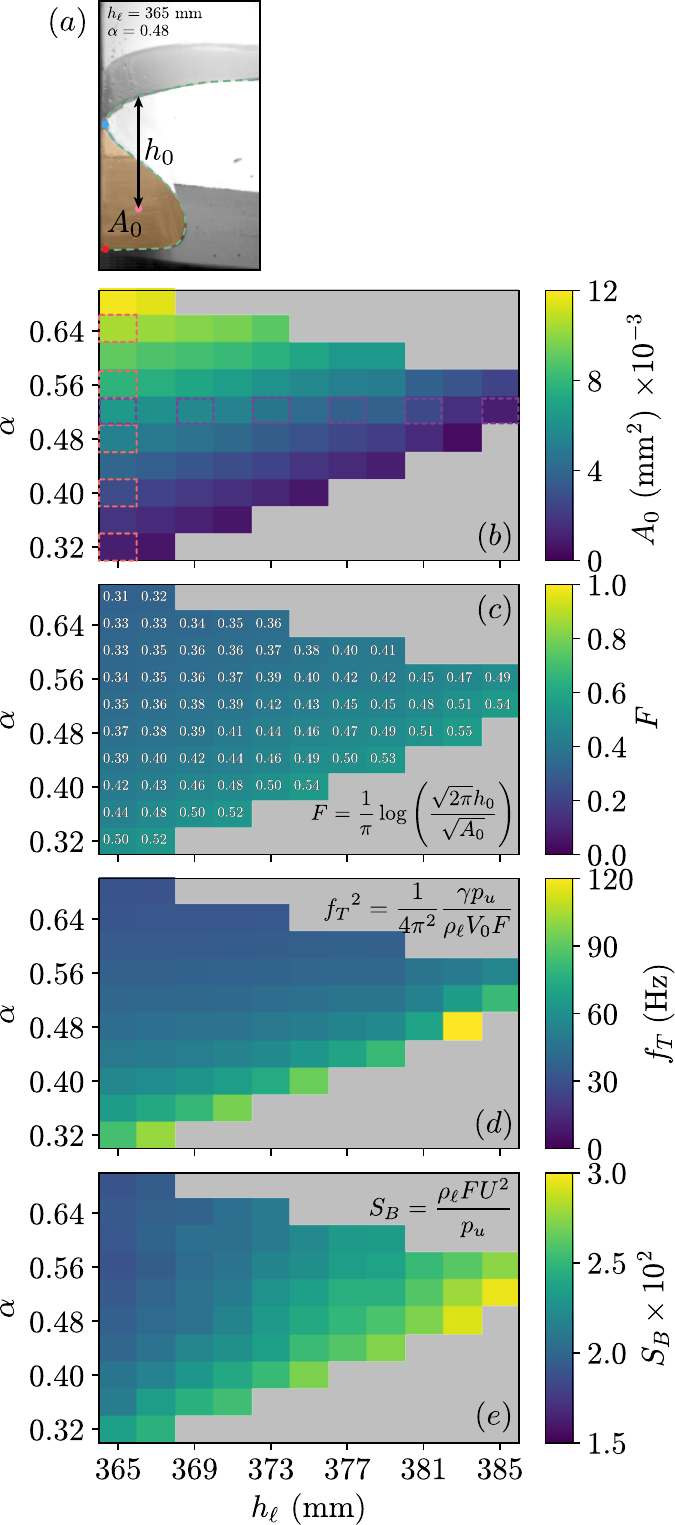}
\caption{(a) Free-surface detection (green dashed line) upon impact for $h_\ell=\qty{365}{\mm}$ and $\alpha = 0.48$. The pink solid circle represents the location of the centroid. The initial volume $A_0$ is calculated by integrating the free-surface from the wave trough (red solid circle) to the wave crest (blue solid circle). (b) $A_0$ in phase space. (c) The function $F$ in phase space. (d) The frequency of oscillation of the gas pocket as estimated by \eref{eq:topliss}. (e) $S_B$ in phase space as estimated by \eref{eq:bagnold}. In panels (b) to (e), the colors and numbers represent the magnitude of their respective quantity as indicated by the colorbar. In (b) the highlighted wave states with dashed rectangles correspond to the wave states shown in \fref{fig:solitonsandhs}.}
\label{fig:bagnold}
\end{figure}

\subsection{The behavior of $p_{\max}$, $p_{\min}$ and $f_{\max}$ in phase space. }
\label{sec:volume}

Firstly, we turn to the calculation of \cite{topliss1992} of the natural frequency of a cylindrical gas cavity $f_T$ that oscillates onto a solid wall. The calculation is based on potential theory and the method of images (surface tension is not included). The solution corresponding to an oscillation pocket near the free-surface is $(2 \pi f_T )^2 = 2\gamma p_0 / (\pi R_0^2 \rho_\ell F )$, where $\gamma=1.402$ is the adiabatic constant for air, $R_0$ is the radius of the cylinder, $p_0$ the ambient pressure and $F=(1/ \pi) \log(2h_0/R_0)$ is a dimensionless geometrical correction that depends on $R_0$ and the distance from the center of the pocket to the free-surface $h_0$. By setting $p_0=p_u$ and $R_0^2=2 A_0/ \pi$, where $A_0$ is the initial area of entrapped gas, we can estimate the frequency of oscillation in our experiments as

\begin{equation}
f_T^2 = \frac{1}{4 \pi^2}\frac{\gamma p_u}{\rho_\ell A_0 F}.
\label{eq:topliss}
\end{equation}

The estimate of the frequency in \eref{eq:topliss} can be used as will be shown later to elucidate some of the trends of $f_{\max}$ in phase space.

Next, we turn our attention to the well-known Bagnold model which has been extensively used in the literature to describe the dynamics of an oscillating gas pocket \citep{bagnold1939,brosset2013,kolkmanc2007,dias2018,bogaert2018, ibrahim2020}. The model is based on the equation of motion of a one-dimensional piston driven by a pressure difference which is often used as an analogy to the motion of a gas cavity that is entrapped by a breaking wave upon impact. A thorough derivation of the model is not presented here -- for more information about the model, we refer the reader to the reviews of \citet{dias2018} and \citet{ibrahim2020}. The corresponding equation of motion in dimensionless form reads $S_B \ddot{x} = x^{-\gamma}-1$, with the initial conditions $x(t=0)=0$ and $\dot{x}(t=0)=-1$. Here, $x\equiv x(t)$ is the dimensionless time dependent piston motion and $S_B$ is the so-called Bagnold (or Impact) number $S_B=m_{\ell}U_0^2/(p_0 V_0) $, where $m_\ell$ is the liquid mass associated to the piston, $U_0$ is the initial piston velocity and $V_0$ is the initial volume of entrapped gas by the piston. $S_B$ essentially quantifies the ratio between the initial kinetic energy of the fluid and the initial potential energy of the gas. The solution to the model depends on two parameters: the adiabatic constant $\gamma$ and $S_B$. The solution can then be used to obtain the dimensionless pressure $p^{\star} = (x^{\star})^{-\gamma} -1$, which can be expressed in dimensional form via $p^{\star}=(p-p_0)/p_0$. More importantly, the model predicts that an increment in $S_B$, leads to a monotonic increment of both the min. $p^{\star}_{\min}$ and the max. pressure $p^{\star}_{\max}$. As an alternative representation of $S_B$, \citet{bogaert2018} proposed to set the liquid mass as $m_\ell=\rho_\ell V_0 F$ which yields $S_B=\rho_\ell F U_0^2 / p_0$. Note that here, $F$ is the same function as in the estimate of the frequency by \citet{topliss1992}. Similarly as with the frequency in \eref{eq:topliss}, we seek to estimate the $S_B$ from the experimental data. With the estimate of $S_B$ at hand, we can then extract both the max. and min. pressure from the solution and compare these to $p_{\max}$ and $p_{\min}$ in phase space. We set thus $p_0=p_u$ and assume that the initial velocity of the piston is the propagation speed of the soliton, i.e. $U_0=U=\sqrt{g h_{\ell}(1 + \alpha)}$ which yields,

\begin{equation}
S_B = \frac{\rho_\ell F U }{p_u}
\label{eq:bagnold}
\end{equation}

Now, both estimates in \eref{eq:topliss} and \eref{eq:bagnold} require the knowledge of $A_0$ and $F$ which are wave state dependent. Thus, for every wave state we extract both quantities from the high-speed recordings. To calculate $A_0$, we firstly extract the wave shape profile upon impact. The detection algorithm is based on manually adding black markers along the free-surface. These markers are then extracted by binarization, ordered by nearest-neighbor and their coordinates (horizontal and vertical) parameterized by the arc length $s$, i.e. the distance along the wave shape with $s=0$ located at the intersection of the impact wall. Lastly, a cubic-spline is applied to both parameterized coordinates which yields the final wave shape profile. An example of this detection is shown in \fref{fig:bagnold}a. With the wave shape profile at hand, we find $A_0$ by numerically integrating the wave shape profile from the wave trough to the wave crest. In \fref{fig:bagnold}b we show $A_0$ in phase space, where we observe that -- similarly to what \frefs{fig:solitonsandhs}a,c,e show -- $A_0$ increases with $\alpha$ for a given $h_\ell$. Conversely, for fixed $\alpha$, we see that $A_0$ monotonically decreases for increasing $h_\ell$ -- which is the same trend shown in \frefs{fig:solitonsandhs}b,d,f. In short, this also leads to a gradient of $A_0$ albeit now directed towards (low/high) values of ($h_{\ell}$/$\alpha$), which is nearly the ``mirrored'' version of the gradients of $p_{\max}$, $|p_{\min}|$ and $f_{\max}$ in \fref{fig:exp}. Note that essentially $A_0$ is also a measure of the initial entrapped volume of gas which can easily obtained with $V_0 =A_0 W$, where $W=\qty{0.6}{\meter}$ is the width of the wave flume.

\begin{figure}[t!]
\centering
\includegraphics[width = \linewidth]{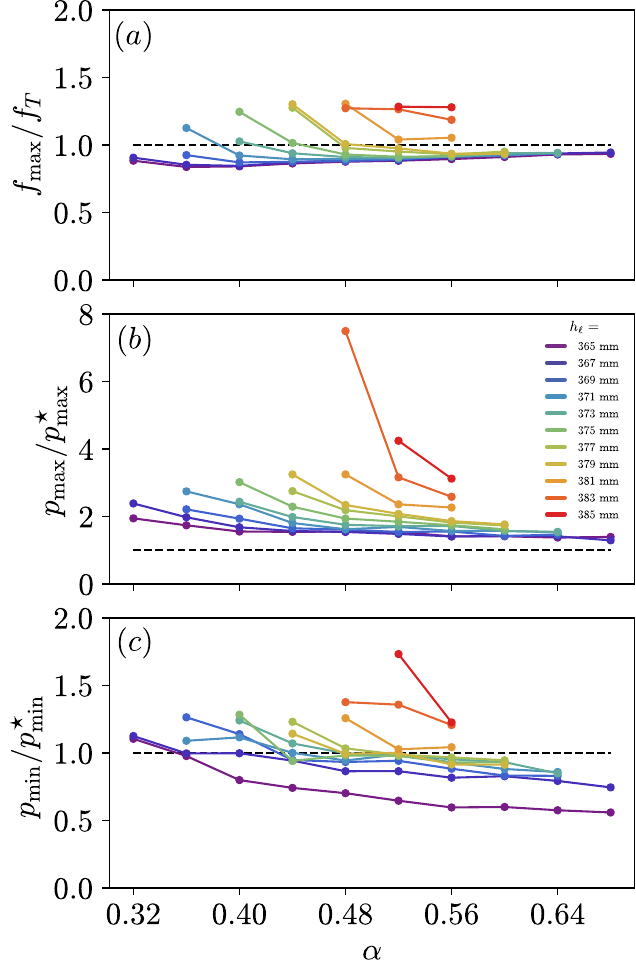}
\caption{Comparison of experimental data to two analytical models. (a) Ratio of experimental dominant frequency of oscillation $f_{\max}$ to the frequency estimate of \cite{topliss1992} as a function of $\alpha$ for various $h_\ell$.  (b) Ratio of experimental max. pressure in gas pocket $p_{\max}$ to max. pressure obtained from the Bagnold model $p_{\max}^{\star}$ as a function of $\alpha$ for various $h_\ell$. (c) Similar as (a) but for the min. pressure in the gas pocket. The values of $S_B$ used to calculate both $p_{\min}^{\star}$ and $p_{\max}^{\star}$ correspond to those shown in \fref{fig:bagnold}e. In all three panels the colors represent different values of $h_\ell$ as shown in the legend of (b).}
\label{fig:comparison}
\end{figure}

Next, we proceed to calculate the function $F$. To do so, we find the centroid of the gas pocket and obtain $h_0$ as shown in the sketch of \fref{fig:bagnold}a. With both $A_0$ and $F$ available, we use \eref{eq:topliss} to estimate the frequency of oscillation of the gas pocket. The estimate is shown in phase space in \fref{fig:bagnold}d, where we obtain a fairly good agreement with the measured frequency that is shown in \fref{fig:exp}c. Qualitatively, both the magnitude and orientation of the gradient is similar and quantitatively the ratio between measured and predicted frequency $f_{\max}/f_T$ is nearly one for most wave states as shown in \fref{fig:comparison}a. Interestingly, the prediction improves when $\alpha$ increases for a given $h_\ell$, i.e. when the wave state is far away from the GP limit. For completeness, we show in \fref{fig:bagnold}b, the function $F$ in phase space. Note that the largest values can be found close to the GP limit and are in the order of $\approx 0.5$, while the smallest value ($0.35$) can be found for the lowest water depth and highest $\alpha$. As the data is in good agreement with \eref{eq:topliss}, we thus attribute the changes of $f_{\max}$ in phase space to the inverse of the factor $A_0 F$, which in itself can also be thought of as an effective initial volume of entrapped gas, i.e. the initial volume of entrapped gas plus a geometric correction due the shape of the pocket via $F$.

In \fref{fig:bagnold}e, we show now $S_B$ in phase space as obtained by \eref{eq:bagnold}, where again observe a well-defined gradient that is oriented towards larger $\alpha$ and $h_\ell$ -- very similar to the gradient of $p_{\max}$ shown in \fref{fig:exp}a. In the context of the Bagnold model, the max. pressure in the gas pocket is an increasing monotonic function of $S_B$. This would suggest then, that when increasing $\alpha$ for a fixed $h_\ell$, the measured max. pressure should decrease -- which is precisely what we experimentally measure in \fref{fig:exp}a. Conversely, when $\alpha$ is fixed in \fref{fig:bagnold}c, we see an increment of $S_B$ for increasing $h_\ell$. And this on the contrary, should lead to an increment of the max. pressure which is what we also obtain experimentally in \fref{fig:exp}a. Note that since the min. pressure in the Bagnold model monotonically increases with $S_B$, a similar argument can be made for the min. pressure shown in \fref{fig:exp}b.   

While this qualitative argument is useful to explain the trends of both $p_{\max}$ and $p_{\min}$ in \fref{fig:exp}, we note that quantitatively, the Bagnold model via \eref{eq:bagnold} underpredicts the value of both max. and min. pressures as show in \fref{fig:comparison}b,c. For the case of $p_{\max}$, we find that for a given $h_\ell$, the prediction improves for large $\alpha$ –– similar to what is found for the frequency of oscillation. We find that at best, the Bagnold model under predicts the measured max. pressure by only a factor of $\approx 1.5$ for the lowest $h_{\ell}=\qty{365}{\milli \meter}$. Conversely, the worst prediction is found for the two largest $h_\ell$ ($\qty{383}{\mm}$ and $\qty{385}{\mm}$), where the Bagnold estimate differs from the measured data by $\approx 4$ and $\approx 8$, respectively. Interestingly, with respect to the prediction of $p_{\min}$, the Bagnold model overall provides better predictions as compared to $p_{\max}$, except for $h_\ell=365 \ \text{mm}$ and the two largest $h_\ell$. Here, we also find that the prediction improves when $\alpha$ increases for a given $h_\ell$. We note that the differences between measured and predicted pressures could also be likely due to the nominal value of $S_B$ that is associated to every wave state. Indeed, given this choice of $S_B$ via \eref{eq:bagnold}, we have assumed that the equivalent radius can be replaced in terms of $A_0$, which is essentially equivalent to assuming that the gas pocket can be approximated as half cylinder as it was done in the calculation of \cite{topliss1992}. Similarly, we have assumed here that the initial piston velocity $U_0$ is equal to the propagation speed of the soliton $U$ which is likely underestimated. In spite of this, we find it remarkable that a simple one-dimensional model is able to predict to first order, both the magnitude of the max. and min. pressures for such a diverse family of gas pockets. Rather than performing a thorough investigation behind these discrepancies -- which is outside of the scope of the current work –– we would like to emphasize that the Bagnold model is used here simply to elucidate the trends shown in \fref{fig:exp}a,b. In this spirit, we attribute thus the behavior of the gradients of $p_{\max}$ and $p_{\min}$ in phase space to an equivalent gradient of $S_B$.

\section{Results: Machine Learning}
\label{chap:results_ml}

For the training of both the MLP and cLSTM models, we adopt the 5-fold cross-validation approach. The data is split in such a way that all the three repetitions for a certain wave state end up either in the training or the validation dataset. When comparing the results of the ML models to the measurements, we utilize the $R^2$ scores. For a total of $m$ samples, the $R^2$ is calculated as 

\begin{equation}
{R^2}\left( {y,\hat y} \right) = 1 - \frac{{\sum\nolimits_{i = 1}^m {{{\left( {{y_i} - {{\hat y}_i}} \right)}^2}} }}{{\sum\nolimits_{i = 1}^m {{{\left( {{y_i} - \bar y} \right)}^2}} }},
\label{eq:r2_score}
\end{equation}

\noindent where ${{\hat y}_i}$ is the predicted value of the $i$-th sample, ${y_i}$ is the corresponding true value and $\bar y$  is its mean, i.e. $= \frac{1}{m}\sum\nolimits_{i = 1}^m {{y_i}}$. The $R^2$ score represents the proportion of variance (of $y$) that has been captured by the independent variables in the model. Moreover, it indicates the goodness of the fit and therefore can be used as a measure of how well unseen samples are likely to be predicted by the model through the proportion of explained variance \cite{scikit-learn}. The highest possible score is $1$. However, its value can also be negative -- as the model in principle can be arbitrarily worse. In general, when the true $y$ is non-constant, a constant model that always predicts the average $y$ disregarding the input features would get a score of $0$.

\begin{figure*}[ht]
\centering
\includegraphics[width = 0.9\linewidth]{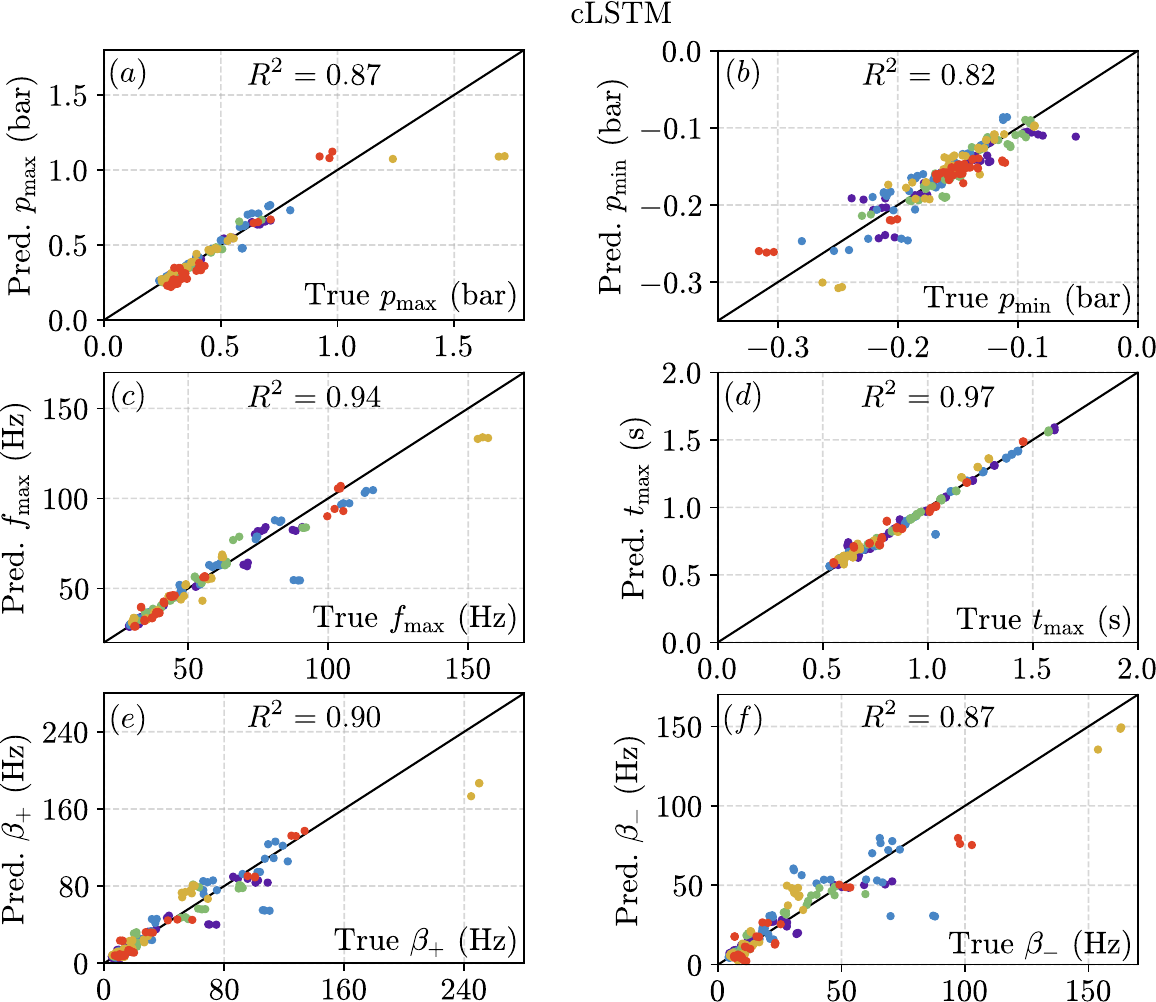}
\caption{True and predicted output quantities from the cLSTM model on the validation datasets from the five-fold cross-validation. The results from each fold are shown in a separate color.}
\label{fig:ConvLSTMscatters}
\end{figure*}

\fref{fig:ConvLSTMscatters} shows the comparison between the predicted values of the output quantities by the cLSTM model and the true values. Here, the data is obtained by concatenating all the validation datasets from the 5-folds. In \fref{fig:ConvLSTMscatters}, we add the corresponding $R^2$ scores for each output and highlight each fold with a different color. We find that overall, the cLSTM model performs well when predicting all the outputs as evidenced by how close the markers are to the identity line and by the corresponding $R^2$ scores. However, we note that the cLSTM underpredicts some of the values. For instance, we find that some of the large errors in $p_{\max}$, $p_{\min}$, $f_{\max}$ and $\beta_+$ are contained within the same fold (see yellow markers in \fref{fig:ConvLSTMscatters}). This fold includes some edge cases in the validation dataset, which might explain the behavior in the scatter plots. The three largest measured values observed in the scatter plots for $p_{\max}$, $f_{\max}$, $\beta_+$ and $\beta_-$ belong to the same wave state ($\alpha=0.48$, $h_\ell=\qty{383}{\milli \meter}$). As mentioned in \sref{sec:outputs_phase_space}, this wave state lies very close to the GP limit where we expect a larger variability of the impact pressures. Indeed, the measured $p_{\max}$ values from the three repetitions are $\qty{1.716}{\bar}, \qty{1.691}{\bar}$ and $\qty{1.237}{\bar}$ which show a large variance, i.e. the last repetition produces a value $27\%$ smaller than the first two (see \fref{fig:avg_output}a). The cLSTM model seems to consistently regress towards the small value for this wave state. In general however, we find that the performance of the cLSTM models is consistent among the different folds.

\begin{table}[h]
    \centering
    \begin{tabular}{l|c|c|c|c|c|c|c|c|c|c}

     & $p_{max}$ & $p_{\min}$ & $t_{\max}$ & $f_{\max}$ & $\beta_+$ & $\beta_-$\\
     \hline
     MLP & 0.93 & 0.94 & 1.0 & 0.98 & 0.96 & 0.94 \\
     cLSTM & 0.87 & 0.82 & 0.97 & 0.94 & 0.90 & 0.87\\     
    \end{tabular}
    \caption{$R^2$ scores for the six output quantities from the MLP and cLSTM models.}
    \label{tab:R2scores}
\end{table}

\begin{figure*}[t!]
    \centering
    \includegraphics[width = \linewidth]{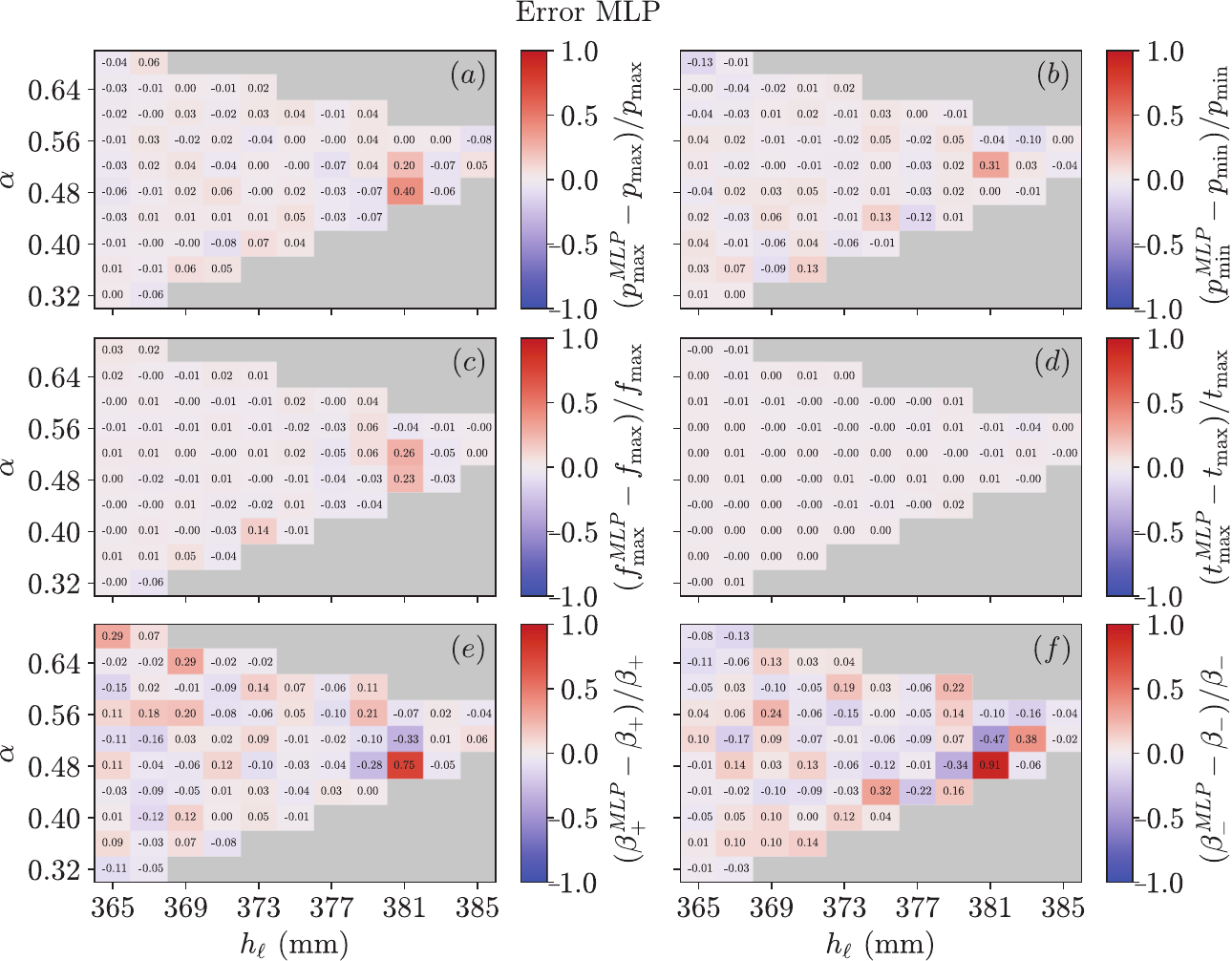}
    \caption{Error of the MLP model in phase space for all outputs. (a) $p_{\max}$, (b) $p_{\min}$, (c) $f_{\max}$, (d) $t_{\max}$, (e) $\beta_{+}$, and (f) $\beta_{-}$. In every panel, the color represents the magnitude of that quantity as described by the corresponding colorbar.}
    \label{fig:err_MLP}
\end{figure*}

\begin{figure*}[t!]
    \centering
    \includegraphics[width = \linewidth]{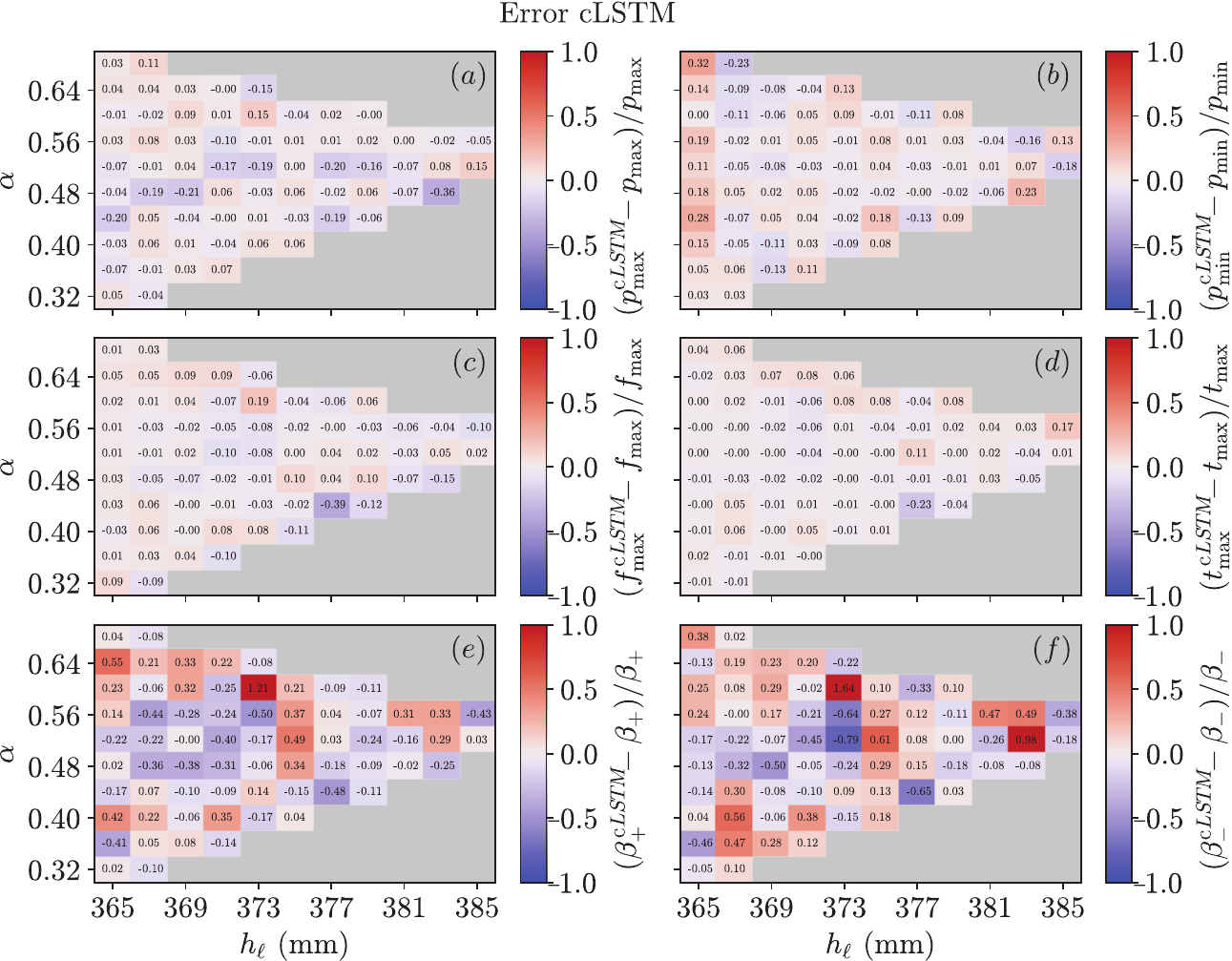}
    \caption{Error of the cLSTM model in phase space for all outputs. (a) $p_{\max}$, (b) $p_{\min}$, (c) $f_{\max}$, (d) $t_{\max}$, (e)  $\beta_{+}$, and (f) $\beta_{-}$. In every panel, the color represents the magnitude of that quantity as described by the corresponding colorbar.}
    \label{fig:err_cLSTM}
\end{figure*}

In order to further reveal the performance of the ML models, we show in Figs. \ref{fig:err_MLP} and \ref{fig:err_cLSTM} the relative error in phase space associated to both MLP and cLSTM models, respectively. The relative error is the difference (in percentage) between the prediction by each model and the experimental data shown in \fref{fig:exp}. For completeness, the actual predictions from both models in phase space are also included in \sref{sec:predMLmodels}. From both models, we find that the relative errors are generally larger for $\beta_{+}$ and  $\beta_{-}$ as compared to the other output variables. For these two output quantities, the models result in over- and under-predictions, and the errors are spread over the whole wave state domain. The relative errors are larger with the cLSTM model as compared to the MLP model. As curve fitting is applied to the pressure signals in order to obtain the decay rates (see \fref{fig:pressure_sensors}b), we note that this might introduce inconsistent values depending on the fitting.

As a summary, \tref{tab:R2scores} lists the $R^2$ scores calculated by considering the performance of both models on the concatenated validation datasets from the 5-folds. Compared to the MLP model, the cLSTM model performs worse resulting in lower $R^2$ scores for all the output quantities. However, we note that the regression task of the cLSTM model is considerably more complex as compared to that of the MLP model. This is evident when one considers their inputs (two high-speed camera snapshots) and model architectures. While the MLP model is trained to approximate the response surfaces shown in Fig. \ref{fig:exp}, the cLSTM model is trained to learn contextual information from the high-definition images of the breaking waves in order to decode this information later to predict the output quantities. Furthermore, the cLSTM model has more trainable parameters than the MLP model, and as such, it is more susceptible to over-fitting. 

\begin{figure*}[ht]
    \centering
    \includegraphics[width = 0.75\linewidth]{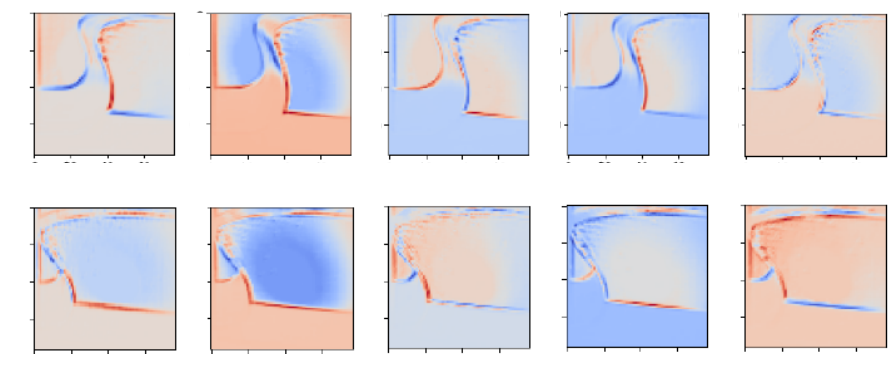}
    \caption{Some filter activations from the input-to-state convolutions in the first ConvLSTM layer}
    \label{fig:activations_cLSTM}
\end{figure*}

Finally, \fref{fig:activations_cLSTM} shows ten selected filter activations from the input-to-state convolutions in the first ConvLSTM layer. The larger values in the activations have darker red colors whereas the smaller values have lighter blue colors. These filters seem to learn different but relevant features in the images. Some filters focus more on the curvature of the crest while others learn the shape of the beach the wave is propagating over. Moreover, the location of the vertical wall is captured by some filters. The strong background illumination present in all experiments seems to also somewhat affect the output of the filters. In general, these results show that the cLSTM model captures relevant features from the data in its regression task and is not merely guessing or memorizing the outputs.

Considering all the results presented above, the following ideas are proposed to potentially improve the performance of the cLSTM model. For some of the output quantities (e.g. $p_{\max}$,  $p_{\min}$, $f_{\max}$, $\beta_{+}$, and $\beta_{-}$), limited number of wave states are available in the dataset in the vicinity of their largest or smallest measured values,  see \fref{fig:ConvLSTMscatters}. Increasing the number of wave states in these regions by conducting additional experiments might be beneficial. Furthermore, the background illumination in the images of the breaking waves can be filtered out during the data preprocessing. To improve the performance of the cLSTM model for predicting decay rates, it may be beneficial to include snapshots taken after the impact in the model's input, as it currently only incorporates snapshots from before the impact.

\section{Conclusions}
\label{chap:conclusions}

We perform wave impact experiments and use this data to train two machine learning models that predict the dynamics of an oscillating gas pocket. This pocket is generated when a breaking wave entraps a certain amount of air in the vicinity of a solid wall. The data consists of a family of breaking waves, which in turn leads to a family of gas pockets. The breaking waves are generated in a flume tank via the interaction of a solitary wave and a beach. We show that the wave generation can be well described in terms of two control parameters. Namely, the ratio of wave amplitude to water depth $\alpha=A/ h_\ell$ and the water depth $h_\ell$. By exploring different pairs of $\alpha$ and $h_\ell$, we are able to define a region wherein breaking is guaranteed, and thus where gas pockets can form. We call this region ``the phase space of wave generation'', where every pair ($h_\ell$, $\alpha$) leads to a unique ``wave state'' that is used to train the ML models. This property is evidenced as every wave state essentially leads to a different GWS upon impact and a different propagation speed of the soliton.

We describe the dynamics of the gas pocket by selecting six output scalars which are averaged within the gas pocket. We observe the formation of well-defined gradients in phase space for all output scalars. In particular,  the gradients of $p_{\max}$, $p_{\min}$, $f_{\max}$ and both decay rates are oriented towards large $\alpha$, low $h_\ell$ and are maximized near the so-called GP limit -- which is characterized by wave states that generate small gas pockets. The trends of $t_{\max}$ can be simply attributed to the propagation speed of the soliton $U$ while the trends of both $\beta_-$ and $\beta_+$ can be explained by a simple geometrical argument derived from $p_{\max},p_{\min}$ and $f_{\max}$. In order to elucidate the trends from these last three gradients however, we measure the initial entrapped volume $V_0$. Here, we find also a well-defined gradient that -- as opposed to $p_{\max}$ -- is directed towards larger $\alpha$ and smaller $h_\ell$.

Furthermore, we find a good agreement between the measured frequency of oscillation $f_{\max}$ and the calculation of \cite{topliss1992}. This suggests that the behavior of $f_{\max}$ in phase space is not solely dependent on $V_0$, but to the combined effect of $V_0$ plus a geometric correction due to the wave shape and the hydrostatic pressure via the function $F$. In terms of the impact pressures, we attribute the behavior of both $p_{\max}$ and $p_{\min}$ (and thus of their respective gradients in phase space) not only to the initial volume $V_0$ but to its combined effect with the wave kinematics via the Bagnold number $S_B$.

Regarding the ML predictions, we find that both the cLSTM and MLP models are able to qualitatively reproduce the gradients for all output variables. Quantitatively, for the MLP model, we find that the errors in $p_{\max}$, $p_{\min}$, and $f_{\max}$ are generally less than $ <10 \%$. Here, the larger errors are found close the GP limit (small gas pockets). Regarding the decay rates, the model yields errors that are \magenta{sometimes} larger than $10 \%$, \magenta{and there is no apparent trend in the error of the model with respect to the wave state}. Similar conclusions can be drawn for the cLSTM model, although the model in general yields larger errors than the MLP. In particular, we observe $p_{\max}$, $p_{\min}$ and $f_{\max}$ errors of less than $20 \%$. While the cLSTM model qualitatively describes the gradients of both decay rates, we sometimes observe large errors ($> 20 \%$) with \magenta{no visible trend with respect to the wave state}. Nonetheless, we find it remarkable that a ML model that simply takes two snapshots as input can yield errors of the max.  and min. impact pressures in the gas pocket that are of the same magnitude as those given by the Bagnold model which is vastly used in the literature.

\magenta{It should be noted that our model is trained with laboratory data where we have control over the fluid properties, the water depth, the WM motion and the bathymetry. As a result, the breaking waves are smooth and the free-surface is well-defined. In contrast, when considering field applications at full scale, this is unlikely to occur and thus the application of our model to that scenario remains challenging. We believe however that our work can still be used as a benchmark for the development of more advanced ML models. }

Furthermore, the work presented here is focused on the dynamics of entrapped air during a wave impact. We believe that a similar methodology can also be applied and extended so as to include additional components. For instance, these models could focus on additional ELPs \magenta{from the pressure map} -- such as direct impacts and running jets; additional types of wave impacts -- such as flip-through and other breakers (spilling, surging, plunging); different bathymetries; and even additional physical phenomena -- such as gas and liquid compressibility, gas-to-liquid density ratio effects, surface tension, phase transition and even fluid-structure interaction. \magenta{Additionally, a potentially promising application of these models could be ``screening'', where wave elevation data is typically used to detect extreme events that lead to large amplitude wave impacts. We note however that some of the examples presented above have inherent statistical distributions that are not accounted for in the approach presented here. Instead, one could turn to uncertainty-aware machine learning models and use those type of models in future work.} 

\section{Acknowledgements}
We would like to thank L. Brosset, D. van der Meer for various stimulating discussions when preparing the manuscript. We would also like to dedicate this work to the memory of L. Kimmoun (IRPHE, Aix-Marseille Universit\'e) who sadly passed away on early March 2023. The wave generation in this work could only be possible due to her contribution -- she will be greatly missed. Finally, we would like to acknowledge the technical support of E. Abrahamse, A. Kasdjo and E. Seves.

\section{Bibliography}
\bibliographystyle{apalike}
\bibliography{MLSolitons}

\begin{thebibliography}{}

\bibitem[Bagnold, 1939]{bagnold1939}
Bagnold, R.~A. (1939).
\newblock Interim {{Report On Wave-Pressure Research}}.
\newblock {\em J. Inst. Civ. Eng.}, 12:202--226.

\bibitem[Bogaert, 2018]{bogaert2018}
Bogaert, H. (2018).
\newblock {\em An Experimental Investigation of Sloshing Impact Physics in
  Membrane {{LNG}} Tanks on Floating Structure}.
\newblock PhD thesis, Delft University of Technology.

\bibitem[Boussinesq, 1871]{boussinesq1871}
Boussinesq, J. (1871).
\newblock {Th{\'e}orie de l'intumescence liquide appel{\'e}e onde solitaire ou
  de translation se propageant dans un canal rectangulaire}.
\newblock {\em J. Math.Pures Appl.}, 72:755--759.

\bibitem[Bredmose et~al., 2015]{bredmose2015}
Bredmose, H., Bullock, G.~N., and Hogg, A.~J. (2015).
\newblock Violent breaking wave impacts. {{Part}} 3. {{Effects}} of scale and
  aeration.
\newblock {\em J. Fluid Mech.}, 765:82--113.

\bibitem[Brosset et~al., 2013]{brosset2013}
Brosset, L., Ghidaglia, J.-M., Guilcher, P.-M., and Tarnec, L.~L. (2013).
\newblock Generalized {{Bagnold Model}}.
\newblock In {\em Proc. of the 23rd {{Int}}. {{Offshore}} and {{Polar Eng}}.
  {{Conf}}. ({{ISOPE}})}.

\bibitem[Buscombe and Carini, 2019]{buscombe2019}
Buscombe, D. and Carini, R.~J. (2019).
\newblock A {{Data-Driven Approach}} to {{Classifying Wave Breaking}} in
  {{Infrared Imagery}}.
\newblock {\em J. Remote Sens}, 11:859.

\bibitem[Chan, 1994]{chan1994}
Chan, E.-S. (1994).
\newblock Mechanics of deep water plunging-wave impacts on vertical structures.
\newblock {\em Coast. Eng.}, 22:115--133.

\bibitem[Cuomo et~al., 2011]{cuomo2011}
Cuomo, G., Piscopia, R., and Allsop, W. (2011).
\newblock Evaluation of wave impact loads on caisson breakwaters based on joint
  probability of impact maxima and rise times.
\newblock {\em Coast. Eng.}, 58:9--27.

\bibitem[Dias and Ghidaglia, 2018]{dias2018}
Dias, F. and Ghidaglia, J.-M. (2018).
\newblock Slamming: {{Recent Progress}} in the {{Evaluation}} of {{Impact
  Pressures}}.
\newblock {\em Annu. Rev. Fluid Mech.}, 50:243--273.

\bibitem[Duong et~al., 2023]{duong2023}
Duong, N.~T., Tran, K.~Q., Luu, L.~X., and Tran, L.~H. (2023).
\newblock Prediction of breaking wave height by using artificial neural
  network-based approach.
\newblock {\em Ocean Eng.}, 182:102177.

\bibitem[Eadi~Stringari et~al., 2021]{eadi2021}
Eadi~Stringari, C., Veras~Guimar{\~a}es, P., Filipot, J.-F., Leckler, F., and
  Duarte, R. (2021).
\newblock Deep neural networks for active wave breaking classification.
\newblock {\em Sci. Rep.}, 11:3604.

\bibitem[Erinin et~al., 2023]{erinin2023}
Erinin, M.~A., Liu, C., Liu, X., Mostert, W., Deike, L., and Duncan, J.~H.
  (2023).
\newblock The effects of surfactants on plunging breakers.
\newblock {\em J. Fluid Mech.}, 972:R5.

\bibitem[Etienne et~al., 2018]{etienne2018}
Etienne, S., Scolan, Y.-M., and Brosset, L. (2018).
\newblock Numerical {{Study}} of {{Density Ratio Influence}} on {{Global Wave
  Shapes Before Impact}}.
\newblock In {\em Proc. of the 37th {{Int}}. {{Conf}}. on {{Ocean Offshore}}
  and {{Artic Eng}}. ({{OMAE}})}.

\bibitem[Ezeta et~al., 2023]{ezeta2023}
Ezeta, R., Kimmoun, L., and Brosset, L. (2023).
\newblock Influence of {{Ullage Pressure}} on {{Wave Impacts Induced}} by
  {{Solitary Waves}} in a {{Flume Tank}}.
\newblock In {\em Proc. of the 33rd {{Int}}. {{Offshore}} and {{Polar Eng}}.
  {{Conf}}. ({{ISOPE}})}.

\bibitem[Faltinsen et~al., 2004]{faltinsen2004}
Faltinsen, O.~M., Landrini, M., and Greco, M. (2004).
\newblock Slamming in marine applications.
\newblock {\em J. Eng. Math.}, 48:187--217.

\bibitem[Faltinsen and Timokha, 2009]{faltinsen2009}
Faltinsen, O.~M. and Timokha, A.~N. (2009).
\newblock {\em Sloshing}.
\newblock Cambridge University Press.

\bibitem[Fenton, 1972]{fenton1972}
Fenton, J. (1972).
\newblock A ninth-order solution for the solitary wave.
\newblock {\em J. Fluid Mech.}, 53:257--271.

\bibitem[Fortin et~al., 2020]{fortin2020}
Fortin, S., Etienne, S., B{\'e}guin, C., Pelletier, D., and Brosset, L. (2020).
\newblock Numerical {{Study}} of the {{Influence}} of {{Weber}} and {{Reynolds
  Numbers}} on the {{Development}} of {{Kelvin-Helmholtz Instability}}.
\newblock {\em Int. J. Offshore Polar Eng.}, 30:129--140.

\bibitem[Goring, 1978]{goring1978}
Goring, D.~G. (1978).
\newblock {\em Tsunamis -- the Propagation of Long Waves onto a Shelf}.
\newblock PhD thesis, California Institute of Technology.

\bibitem[Grimshaw, 1971]{grimshaw1971}
Grimshaw, R. (1971).
\newblock The solitary wave in water of variable depth. {{Part}} 2.
\newblock {\em J. Fluid Mech.}, 46:611--622.

\bibitem[Hofland et~al., 2011]{hofland2011}
Hofland, B., Kaminski, M., and Wolters, G. (2011).
\newblock Large {{Scale Wave Impacts}} on a vertical wall.
\newblock In {\em Coast. {{Eng}}. {{Proc}}.}, volume~32.

\bibitem[Ibrahim, 2020]{ibrahim2020}
Ibrahim, R.~A. (2020).
\newblock Assessment of breaking waves and liquid sloshing impact.
\newblock {\em Nonlinear Dyn.}, 100:1837--1925.

\bibitem[Kapsenberg, 2011]{kapsenberg2011}
Kapsenberg, G.~K. (2011).
\newblock Slamming of ships: Where are we now?
\newblock {\em Phil. Trans. R. Soc. A.}, 369:2892--2919.

\bibitem[Karimi et~al., 2015]{karimi2015}
Karimi, M., Brosset, L., Ghidaglia, J.-M., and Kaminski, M. (2015).
\newblock Effect of ullage gas on sloshing, {{Part I}}: {{Global}} effects of
  gas--liquid density ratio.
\newblock {\em Eur. J. Mech. B Fluids}, 53:213--228.

\bibitem[Katell and Eric, 2002]{guizien2002}
Katell, {\relax Guizien}. and Eric, B. (2002).
\newblock Accuracy of solitary wave generation by a piston wave maker.
\newblock {\em J. Hydraul. Res.}, 40:321--331.

\bibitem[Kim and Lee, 2024]{kim2024}
Kim, T. and Lee, W.-D. (2024).
\newblock Prediction of wave runup on beaches using interpretable machine
  learning.
\newblock {\em Ocean Eng.}, 297:116918.

\bibitem[Kimmoun et~al., 2009]{kimmoun2009}
Kimmoun, O., Scolan, Y.~M., and Malenica, {\v S}. (2009).
\newblock Fluid {{Structure Interactions Occuring At}} a {{Flexible Vertical
  Wall Impacted By}} a {{Breaking Wave}}.
\newblock In {\em Proc. of the 19th {{Int}}. {{Offshore}} and {{Polar Eng}}.
  {{Conf}}. ({{ISOPE}})}.

\bibitem[Kolkman and Jongeling, 2007]{kolkmanc2007}
Kolkman, P.~A. and Jongeling, T. H.~G. (2007).
\newblock {\em Dynamic Behaviour of Hydraulic Structures}.
\newblock WL{\textbar} Delft Hydraulics publication.

\bibitem[Lafeber et~al., 2012]{lafeber2012}
Lafeber, W., Bogaert, H., and Brosset, L. (2012).
\newblock Elementary {{Loading Processes}} ({{ELP}}) {{Involved In Breaking
  Wave Impacts}}: {{Findings From}} the {{Sloshel Project}}.
\newblock In {\em Proc. of the 22nd {{Int}}. {{Offshore}} and {{Polar Eng}}.
  {{Conf}}. ({{ISOPE}})}.

\bibitem[{Lay-Ekuakille} et~al., 2021]{lay2021}
{Lay-Ekuakille}, A., Djungha~Okitadiowo, J.~P., Di~Luccio, D., Palmisano, M.,
  Budillon, G., Benassai, G., and Maggi, S. (2021).
\newblock Image {{Sensors}} for {{Wave Monitoring}} in {{Shore Protection}}:
  {{Characterization}} through a {{Machine Learning Algorithm}}.
\newblock {\em Sensors}, 21:4203.

\bibitem[Liu et~al., 2024]{liu2024}
Liu, Y., Eeltink, D., {van den Bremer}, T.~S., and Adcock, T. A.~A. (2024).
\newblock A machine learning architecture for including wave breaking in
  envelope-type wave models.
\newblock {\em Ocean Eng.}, 305:118009.

\bibitem[Lugni et~al., 2006]{lugni2006}
Lugni, C., Brocchini, M., and Faltinsen, O.~M. (2006).
\newblock Wave impact loads: {{The}} role of the flip-through.
\newblock {\em Phys. Fluids}, 18:122101.

\bibitem[Maillard and Brosset, 2009]{maillard2009}
Maillard, S. and Brosset, L. (2009).
\newblock Influence of density ratio between liquid and gas on sloshing model
  test results.
\newblock {\em Int. J. Offshore Polar Eng.}, 19.

\bibitem[Novakovic et~al., 2020]{novakovic2020}
Novakovic, V., Costas, J.~J., Schreier, S., Kimmoun, O., Fernandes, A., Ezeta,
  R., Birvalski, M., and Bogaert, H. (2020).
\newblock Study of {{Global Wave Repeatability}} in the {{New Multiphase Wave
  Lab}} ({{MWL}}).
\newblock In {\em Proc. of the 30th {{Int}}. {{Offshore}} and {{Polar Eng}}.
  {{Conf}}. ({{ISOPE}})}.

\bibitem[Pedregosa et~al., 2011]{scikit-learn}
Pedregosa, F., Varoquaux, G., Gramfort, A., Michel, V., Thirion, B., Grisel,
  O., Blondel, M., Prettenhofer, P., Weiss, R., Dubourg, V., Vanderplas, J.,
  Passos, A., Cournapeau, D., Brucher, M., Perrot, M., and Duchesnay, {\'E}.
  (2011).
\newblock Scikit-learn: {{Machine Learning}} in {{Python}}.
\newblock {\em J. Mach. Learn. Res.}, 12:2825--2830.

\bibitem[Pena and Huang, 2021]{pena2021}
Pena, B. and Huang, L. (2021).
\newblock Wave-{{GAN}}: {{A}} deep learning approach for the prediction of
  nonlinear regular wave loads and run-up on a fixed cylinder.
\newblock {\em Coast. Eng.}, 167:103902.

\bibitem[Peregrine, 2003]{peregrine2003}
Peregrine, D.~H. (2003).
\newblock Water-wave impact on walls.
\newblock {\em Annu. Rev. Fluid Mech.}, 35:23--43.

\bibitem[Rayleigh, 1876]{rayleigh1876}
Rayleigh, L. (1876).
\newblock On waves.
\newblock {\em Phil. Mag.}, 1:257--259.

\bibitem[Saviz~Naeini and Snaiki, 2024]{saviz2024}
Saviz~Naeini, S. and Snaiki, R. (2024).
\newblock A physics-informed machine learning model for time-dependent wave
  runup prediction.
\newblock {\em Ocean Eng.}, 295:116986.

\bibitem[Scharnke et~al., 2023]{scharnke2023}
Scharnke, J., {van Essen}, S.~M., and Seyffert, H.~C. (2023).
\newblock Required test durations for converged short-term wave and impact
  extreme value statistics--{{Part}} 2: {{Deck}} box dataset.
\newblock {\em Mar. Struc.}, 90:103411.

\bibitem[Serre, 1953]{serre1953}
Serre, F. (1953).
\newblock {Contribution {\`a} l'{\'e}tude des {\'e}coulements permanents et
  variables dans les canaux}.
\newblock {\em Houille Blanche}, 39:830--872.

\bibitem[Shi et~al., 2015]{shi2015convolutional}
Shi, X., Chen, Z., Wang, H., Yeung, D.-Y., Wong, W.-k., and WOO, W.-c. (2015).
\newblock Convolutional {{LSTM Network}}: {{A Machine Learning Approach}} for
  {{Precipitation Nowcasting}}.
\newblock In {\em Adv. in {{Neural Inf}}. {{Proc}}. {{Sys}}. ({{NIPS}})}.

\bibitem[Smith et~al., 2023]{smith2023}
Smith, R., Dias, F., Facciolo, G., and Murphy, T.~B. (2023).
\newblock Pre-computation of image features for the classification of dynamic
  properties in breaking waves.
\newblock {\em Eur. J. Remote Sens}, 56:2163707.

\bibitem[Stagonas et~al., 2011]{stagonas2011}
Stagonas, D., Warbrick, D., Muller, G., and Magagna, D. (2011).
\newblock Surface tension effects on energy dissipation by small scale,
  experimental breaking waves.
\newblock {\em Coast. Eng.}, 58:826--836.

\bibitem[Tarwidi et~al., 2023]{tarwidi2023}
Tarwidi, D., Pudjaprasetya, S.~R., Adytia, D., and Apri, M. (2023).
\newblock An optimized {{XGBoost-based}} machine learning method for predicting
  wave run-up on a sloping beach.
\newblock {\em MethodsX}, 10:102119.

\bibitem[Topliss et~al., 1992]{topliss1992}
Topliss, M.~E., Cooker, M.~J., and Peregrine, D.~H. (1992).
\newblock Pressure {{Oscillations During Wave Impacts}} on {{Vertical Walls}}.
\newblock In {\em Proc. of the {{Int}}. {{Conf}}. on {{Coast}}. {{Eng}}.},
  volume~23, pages 1639--1650.

\bibitem[Tu et~al., 2018]{tu2018}
Tu, Y., Cheng, Z., and Muskulus, M. (2018).
\newblock Detection of {{Plunging Breaking Waves Based}} on {{Machine
  Learning}}.
\newblock In {\em Proc. of the 37th {{Int}}. {{Conf}}. on {{Ocean Offshore}}
  and {{Artic Eng}}. ({{OMAE}})}.

\bibitem[van~der Meer, 2022]{vandermeer2022}
van~der Meer, D. (2022).
\newblock Linear stability analysis of a time-divergent slamming flow.
\newblock {\em J. Fluid Mech.}, 934:A4.

\bibitem[{van Essen} and Seyffert, 2023]{vanessen2023}
{van Essen}, S. and Seyffert, H. (2023).
\newblock Finding {{Dangerous Waves}}---{{Review}} of {{Methods}} to {{Obtain
  Wave Impact Design Loads}} for {{Marine Structures}}.
\newblock {\em J. Offshore Mech. Arct. Eng}, 145:060801.

\bibitem[{van Essen} et~al., 2023]{vanessen2023b}
{van Essen}, S.~M., Scharnke, J., and Seyffert, H.~C. (2023).
\newblock Required test durations for converged short-term wave and impact
  extreme value statistics --- {{Part}} 1: {{Ferry}} dataset.
\newblock {\em Mar. Struc.}, 90:103410.

\bibitem[{van Meerkerk} et~al., 2020]{vanmeerkerk2020}
{van Meerkerk}, M., Poelma, C., Hofland, B., and Westerweel, J. (2020).
\newblock Experimental investigation of wave tip variability of impacting
  waves.
\newblock {\em Phys. Fluids}, 32:082110.

\bibitem[Wang and Liu, 2022]{wang2022}
Wang, Y. and Liu, P. L.-F. (2022).
\newblock On finite amplitude solitary waves---{{A}} review and new
  experimental data.
\newblock {\em Phys. Fluids}, 34:101304.

\bibitem[Wu et~al., 2016]{wu2016}
Wu, N.-J., Hsiao, S.-C., Chen, H.-H., and Yang, R.-Y. (2016).
\newblock The study on solitary waves generated by a piston-type wave maker.
\newblock {\em Ocean Eng.}, 117:114--129.

\bibitem[Xiang et~al., 2020]{xiang2020}
Xiang, T., Istrati, D., Yim, S.~C., Buckle, I.~G., and Lomonaco, P. (2020).
\newblock Tsunami {{Loads}} on a {{Representative Coastal Bridge Deck}}:
  {{Experimental Study}} and {{Validation}} of {{Design Equations}}.
\newblock {\em J. Waterway, Port, Coastal, Ocean Eng.}, 146:04020022.

\bibitem[Yun et~al., 2022]{yun2022}
Yun, M., Kim, J., and Do, K. (2022).
\newblock Estimation of {{Wave-Breaking Index}} by {{Learning Nonlinear
  Relation Using Multilayer Neural Network}}.
\newblock {\em J. Mar. Sci. Eng.}, 10:50.

\end{thebibliography}

\appendix

\section{The MLP model}
\label{app:theMLPmodel}

\begin{figure}[ht]
    \centering
    \includegraphics[width = \linewidth]{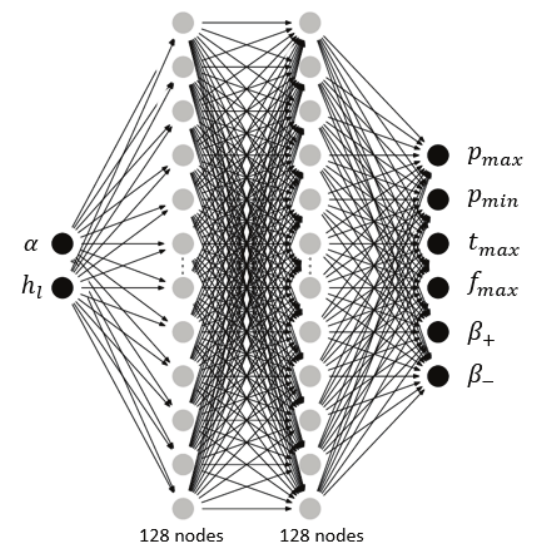}
    \caption{Architecture of the MLP model.}
    \label{fig:mlpDiag}
\end{figure}

For the regression problem where a nonlinear mapping between two inputs and six output quantities (listed in \tref{tab:outputs}) needs to be made, we use a MLP model. Here, we use the machine learning library of \citet{scikit-learn}, where a two-layer MLP model is designed with 128 nodes per layer with ReLU activation function as shown in the sketch of \fref{fig:mlpDiag}. The model is trained for 1000 epochs with a learning rate of $0.001$ using the ADAM optimizer. Both the input and output quantities are normalized by removing the mean and scaling to unit variance. The model has nearly 18K trainable parameters.

\section{Predictions from the Machine Learning Models}
\label{sec:predMLmodels}
\begin{figure*}[ht]
    \centering
    \includegraphics[width = \linewidth]{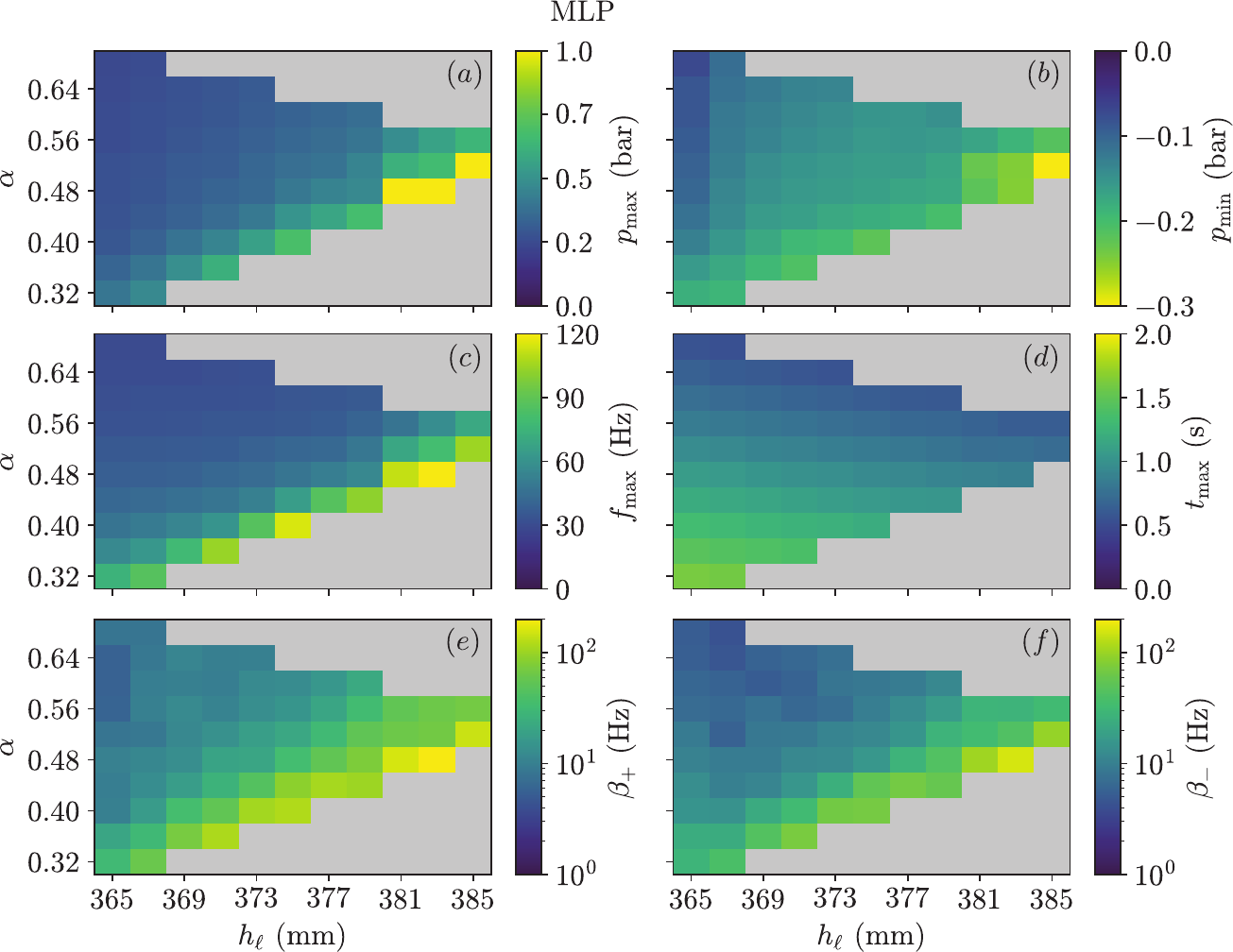}
    \caption{Prediction of the target outputs in phase space from the MLP model. (a) $p_{\max}$, (b) $p_{\min}$, (c) $f_{\max}$, (d) $t_{\max}$, (e) $\beta_{+}$, and (f) $\beta_{-}$. In every panel, the color represents the magnitude of that quantity as described by the corresponding colorbar.}
    \label{fig:MLP}
\end{figure*}

\begin{figure*}[ht]
    \centering
    \includegraphics[width = \linewidth]{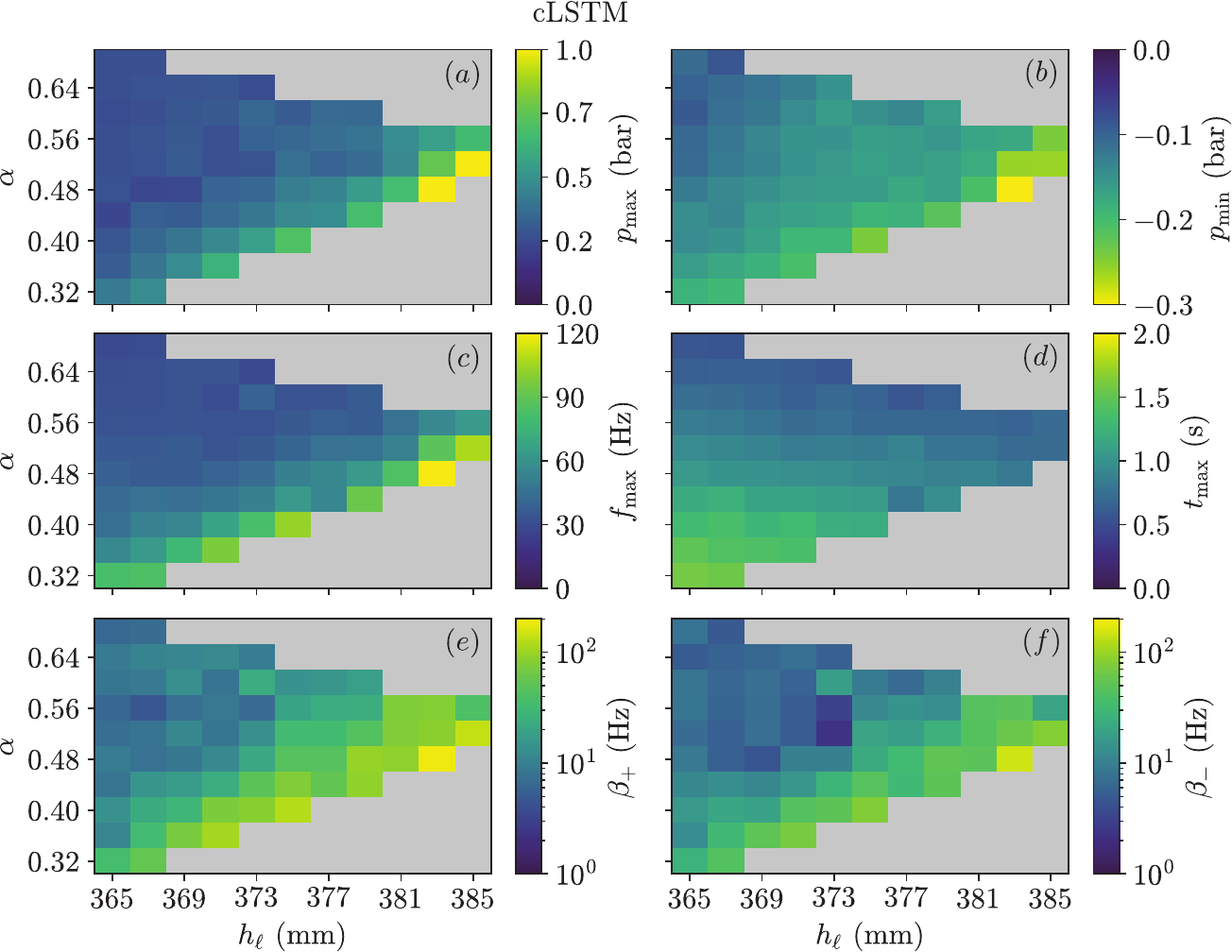}
    \caption{Prediction of the target outputs in phase space from the cLSTM model. (a) $p_{\max}$, (b) $p_{\min}$, (c) $f_{\max}$, (d) $t_{\max}$, (e) $\beta_{+}$, and (f) $\beta_{-}$. In every panel, the color represents the magnitude of that quantity as described by the corresponding colorbar.}
    \label{fig:cLSTM}
\end{figure*}

Figures \ref{fig:MLP} and \ref{fig:cLSTM} show the predictions of the output quantities from the MLP and cLSTM model in phase space, respectively. Note that the models qualitatively capture all the gradients in phase space obtained from the experimental data as shown \fref{fig:exp}.

\printcredits

\end{document}